\documentclass[12pt,prd,preprint,oldfontcommands,nofootinbib]{revtex4-1}
\usepackage{graphicx}
\usepackage{amssymb}
\usepackage{epstopdf}
\usepackage{ mathrsfs }
\usepackage{amsmath}
\usepackage{natbib}
\usepackage[pdftex,bookmarks,linktocpage,plainpages=false,hyperfigures,linkcolor=blue,citecolor=blue,urlcolor=blue]{hyperref} 
\hypersetup{colorlinks=true}

\begin{document}

\title{The Scientific Reach of Multi-Ton Scale Dark Matter Direct Detection Experiments}

\author{Jayden L. Newstead$^{\bf a}$}
\author{Thomas D. Jacques$^{\bf a}$}
\author{Lawrence M. Krauss$^{\bf a, b}$}
\author{James B. Dent$^{\bf c}$}
\author{Francesc Ferrer$^{\bf d}$}

\affiliation{$^{\bf a}$ Department of Physics and School of Earth and Space Exploration, Arizona State University, Tempe, AZ 85287, USA,}

\affiliation{$^{\bf b}$ Research School of Astronomy and Astrophysics, Mt. Stromlo Observatory, Australian National University, Canberra 2614, Australia,}

\affiliation{$^{\bf c}$ Department of Physics, University of Louisiana at Lafayette, Lafayette, LA 70504, USA,}

\affiliation{$^{\bf d}$ Physics Department and McDonnell Center for the Space Sciences, Washington University, St Louis, MO 63130, USA}

\date{\today}

\begin{abstract}

The next generation of large scale WIMP direct detection experiments have the potential to go beyond the discovery phase and reveal detailed information about both the particle physics and astrophysics of dark matter.  We report here on early results arising from the development of a detailed numerical code modeling the proposed DARWIN detector, involving both liquid argon and xenon targets.  We incorporate realistic detector physics, particle physics and astrophysical uncertainties and demonstrate to what extent two targets with similar sensitivities can remove various degeneracies and allow a determination of dark matter cross sections and masses while also probing rough aspects of the dark matter phase space distribution.  
We find that, even assuming dominance of spin-independent scattering, multi-ton scale experiments still have degeneracies  that depend sensitively on the dark matter mass, and on the possibility of isospin violation and inelasticity in interactions.  
We find that these experiments are best able to discriminate dark matter properties for dark matter masses less than around 200 GeV.
 In addition, and somewhat surprisingly, the use of two targets gives only a small improvement (aside from the advantage of different systematics associated with any claimed signal) in the ability to pin down dark matter parameters when compared with one target of larger exposure.
\end{abstract}

\maketitle

\section{Introduction}

Weakly interacting massive particles (WIMPs), among the favored candidates for dark matter, have thus far not been conclusively detected in experiments sensitive to WIMP scattering with nuclei.  A new generation of larger and more diverse detectors is under development, which motivates a consideration of the physics reach of these experiments in order to guide in their design, and also to focus on which uncertainties will be most significant in constraining the conclusions one may derive from any purported detection \cite{Akrami:2010dn,Pato:2010zk,Shan:2011jz,Shan:2011ka,Pato:2011de,Strege:2012kv,Baudis:2012ig,Cerdeno:2013gqa,Arina:2013jya,Billard:2013qya}.  As these detectors become more complex and more expensive, the biggest design effort should reside in ensuring that astrophysical and particle physics degeneracies that will confuse the interpretation of any signal observed by the detectors are reduced as much as possible.  What is required is a realistic, comprehensive numerical tool to model the detectors and the relevant  physics, and one which can be easily modified as design parameters develop and new astrophysical and particle physics constraints evolve.  We have recently set out to complete such a task.

The DARWIN (DARk matter search WIth Noble liquids) project involves a proposed multi-ton detector, based on noble-liquid time projection chamber technology that has been demonstrated with xenon~\cite{Aprile:2012nq} and argon targets~\cite{Benetti:2007cd}. These are complementary targets, since they are well separated in atomic mass, leading to peak sensitivities at different dark matter masses. (For an in-depth description of the DARWIN detector see \cite{Baudis:2010ch,Schumann:2011ts,Baudis:2012bc}).  While the effect of complementarity has been studied for a number of target combinations \cite{Pato:2010zk,Shan:2011jz,Shan:2011ka,Pato:2011de,Baudis:2012ig,Cerdeno:2013gqa}, DARWIN is currently the furthest developed proposal for a direct detection experiment with multiple targets. In this paper we report on the results obtained from the development of a numerical tool that allows a rapid exploration of proposed signals using the most up to date particle physics constraints, astrophysical constraints, and background data, including possible isospin violation, inelastic interactions (in the WIMP sector), different dark matter phase space estimates, and solar neutrino and other detector backgrounds.  We explore degeneracies between different sources of confusion, and point out which areas of experimental and theoretical investigation are likely to be most fruitful if one wants to best exploit co-located detectors containing different noble liquids.

We find that for WIMP masses less than around 200 GeV, the use of two targets can reduce mass and cross section degeneracies and enhance discrimination in the mass-cross section plane, relative to increasing the exposure of either individual target, in agreement with \cite{Pato:2010zk}.

\section{Particle Physics and Astrophysics Inputs}

\subsection{General Formalism}

The primary quantity of interest in direct detection experiments is the differential event rate. In our initial analysis we will focus on WIMPs with spin-independent interactions, in part for simplicity and in part to connect with most of the previous detector development literature, which has focused on this scenario. In a future work we will extend this analysis to include the impact of possible spin dependence  (see for example \cite{Engel:1989ix,Iachello:1990ut,Engel:1991wq,Kamionkowski:1994rm,Jungman1996195}) upon the physics reach of DARWIN and similar detectors.

With respect to the recoil energy $E_R$, the differential rate per nuclei per unit time is
\begin{equation}
\frac{dR}{dE_R} = \frac{\rho_\chi}{m_{\chi}m_N} \int_{|\mathbf{v}| > v_{min}} |\mathbf{v}| f(\mathbf{v}) \frac{d\sigma}{dE_R} d^3\mathbf{v},
\label{rate1}
\end{equation}
where $\rho_\chi$ is the local dark matter density, and $m_{\chi}$, $m_{N}$ are the WIMP and nucleus masses, respectively. The integral averages over the velocity distribution of WIMPs, $f(\mathbf{v})$, weighted by the differential cross section $\frac{d\sigma}{dE_R}$. Kinematically the minimum velocity, $v_{min}$, that can contribute to a recoil of energy $E_R$ is \cite{Pato:2011de}
\begin{equation}
v_{min} = \frac{1}{\sqrt{2 E_R m_N}}\left(\frac{E_R m_N}{\mu_{\chi N}} + \delta \right),
\end{equation} 
where $\mu_{\chi N }$ is the WIMP-nucleus reduced mass and $\delta$ is an inelastic scattering parameter ($\delta=0$ recovers the elastic case). (We note that inelastic scattering is not a property of most WIMP models, but this possibility has been raised \cite{TuckerSmith:2001hy}, and thus we include it here for completeness.) While we are interested in the energy spectrum of the recoils, the full rate can be obtained by integrating this over the range of recoil energies that the detector is sensitive to. The standard approach is to write the cross section in terms of the WIMP-nucleon cross section at zero momentum transfer, $\sigma_0$, and the nuclear form factor, $F^2(E_R)$, 
\begin{equation}
\frac{d\sigma}{dE_R} = \frac{m_N}{2 v^2 \mu_{\chi N }^2} \sigma_0 F^2(E_R).
\end{equation}
The WIMP-nucleon cross section can be written in terms of contributions from neutron and proton scattering, $\sigma_0 = \frac{4 \mu_{\chi N}^2}{\pi} [Z f_p+(A-Z) f_n]^2$, where $A$ and $Z$ are the atomic mass and number of the detector material, $\sigma_{\chi n} = \frac{4 \mu_{\chi n}^2}{\pi} f_{n}^2$ and  $\sigma_{\chi p} = \frac{4 \mu_{\chi p}^2}{\pi} f_{p}^2$ . 
Setting the proton and neutron masses to be equal, an appropriate approximation at the level of accuracy of relevance here,  allows one to write $\sigma_{\chi n}= \left(\frac{f_n}{f_p}\right)^2\sigma_{\chi p}$, such that the factor $\frac{f_n}{f_p}$ neatly incorporates isospin violating interactions. Eq.~\ref{rate1} then becomes
\begin{equation}
\frac{dR}{dE_R} = \frac{\sigma_{\chi p}}{2 m_{\chi} \mu^2_{\chi p } } \left(Z + \frac{f_n}{f_p}\left(A-Z\right)\right)^2\, F^2(E_R) \mathcal{G}(v_{min}),
\label{eqndN}
\end{equation}
where we have defined
\begin{equation}
\mathcal{G}(v_{min}) =  \rho_\chi \int_{|\mathbf v| > v_{min}} \frac{f(\mathbf{v})}{|\mathbf v|}\, d^3\mathbf{v}.
\label{eqnVminInt}
\end{equation}
Using this formalism, the astrophysical and particle physics/nuclear physics inputs are each contained in separate terms, allowing us to examine each in turn.

\subsection{Particle and Nuclear Physics Parameters}

\subsubsection{Isospin and Inelasticity}

\begin{figure}[htp]
\centering
\mbox{
\includegraphics[width=80mm]{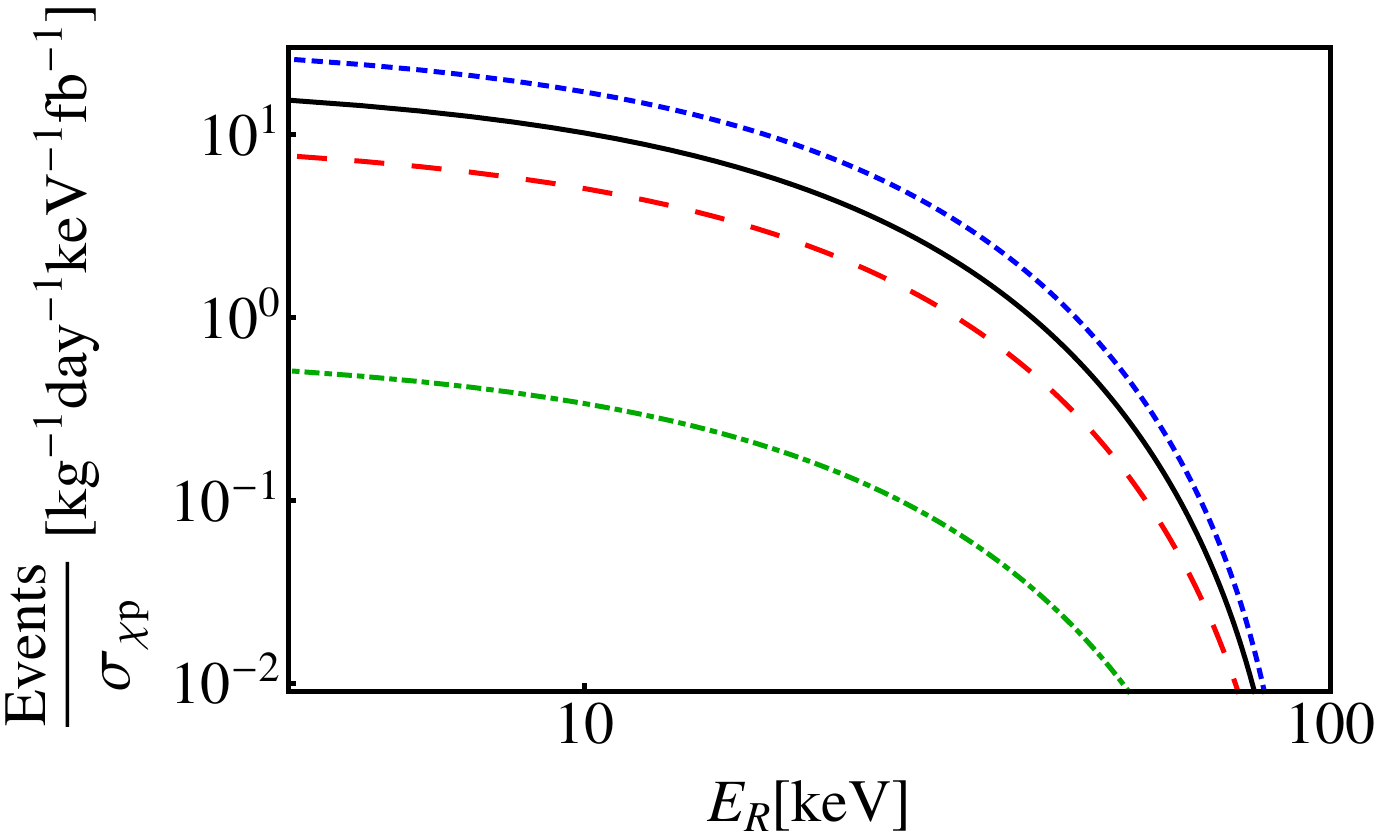}
\includegraphics[width=80mm]{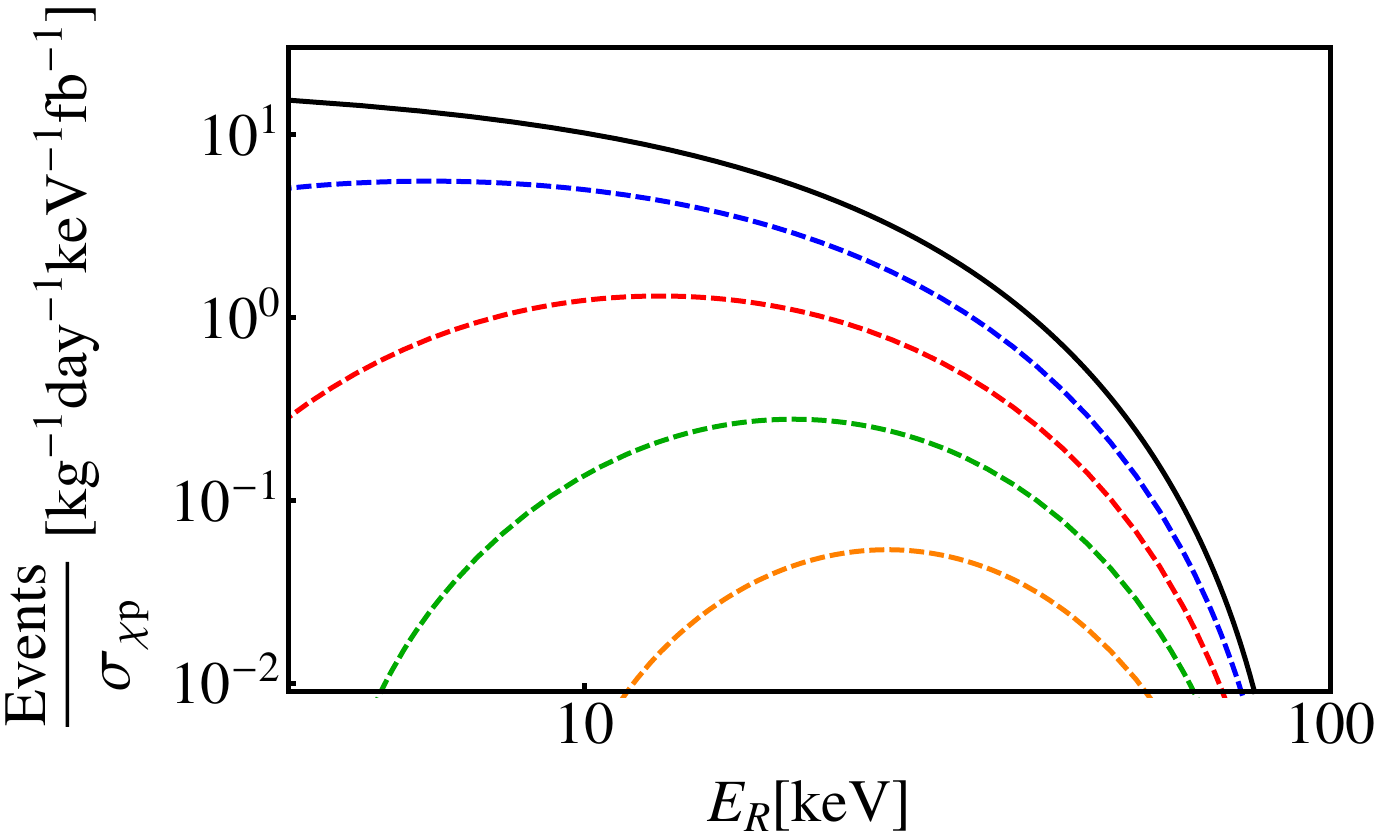}
}
\caption{ The differential event rate per femtobarn of cross section for various values of the isospin violating factor (\textit{left}) and the inelastic parameter (\textit{right}), for a WIMP with $m_\chi = 100$ GeV in a Xenon target, compared to a benchmark WIMP model with the same mass (solid line). A Maxwell-Boltzmann phase-space distribution and the Helm form factor have been assumed (see later sections).
\textit{Left:} From top to bottom, $f_n/f_p = \{1.5,1,0.5,-1\}$. 
\textit{Right:} From top to bottom, $\delta = \{0,25,50,75,100\}$ keV.}
\label{figIEI}
\end{figure}

We have assumed here a simple spin-independent scattering amplitude which means that at low energy the scattering cross section on a nucleus is a simple constant times some product of nuclear charges squared.  While this simplifies the analysis greatly there nevertheless remain two important unknowns related to the specific particle physics parameters of the WIMP sector.  The first involves the WIMP couplings to different quarks, which at low energies get translated into possible isospin violations in the WIMP scattering cross section.  The second involves the (at present, less generic) scenario of excitations in the WIMP sector, which would produce possible inelasticity in the WIMP cross section, parametrized by the quantity $\delta$ mentioned earlier.  
When  the isospin factor is not unity or the inelastic parameter is non-zero, the spectrum is modified, as shown in Fig.~\ref{figIEI}. The isospin factor only affects the magnitude of the recoil rate, while the inelastic parameter severely modifies the shape of the recoil spectrum, as can be seen from Eq.~\ref{eqndN}. The result is that experiments sensitive to the shape of the recoil spectrum are able to determine the value of the inelastic parameter but not the isospin factor, which therefore suffers a degeneracy with the cross section. 

\vspace{25mm}

\subsubsection{Form Factors}
The nuclear form factor encodes the energy dependence of the WIMP-proton cross section, allowing us to derive limits on the cross section at zero momentum transfer. In the lowest order Born approximation, the form factor is the Fourier transform of the nuclear mass distribution. Approximating the nuclei as spherically symmetric we have
\begin{equation}
F(q) = \int _0^{\infty }\rho (r)\frac{\sin (q r)}{q r}4\pi  r^2dr,
\end{equation}
where $q$ is the momentum transfer. The mass distribution of nuclei is not well known, and instead it is generally assumed that the nuclei's mass distribution is approximately the same as its charge-distribution. The most commonly used fits to the charge distribution are the two and the three parameter Fermi distributions (2PF/3PF),
\begin{eqnarray}
\rho_{\tiny{\textrm{2PF}}}(r) = \frac{1}{1+\textrm{exp}(\frac{r-c}{z})}, \\
\rho_{\tiny{\textrm{3PF}}}(r) = \frac{1+w\frac{r^2}{c^2}}{1+\textrm{exp}(\frac{r-c}{z})}, \\
\nonumber
\end{eqnarray}
where the normalization is obtained by requiring $F(q=0)=1$. Unfortunately these distributions do not have analytic Fourier transforms. Instead it is common to use the analytic Helm form factor, obtained by convolving a constant, spherical charge distribution with a `fuzzy' skin. The Helm form factor is given by \cite{Engel1992310}
\begin{equation}
F(q) = 3\frac{\textrm{sin}(q r_n) - q r_n\textrm{cos}(q r_n)}{(q r_n)^3}\exp\left[\frac{-(q s)^2}{2}\right],
\end{equation}
where the skin thickness $s \approx 0.9 $ fm and we use \cite{Lewin199687}
\begin{equation}
r_n^2  = \left( (1.23 A^{1/3}-0.6)^2 +\frac{7}{3}\pi^2 (0.52)^2 - 5 \left(\frac{s}{\rm fm}\right)^2 \right){\rm fm}^2.
\end{equation}

\begin{figure}
\centering
\mbox{
\includegraphics[height=54mm]{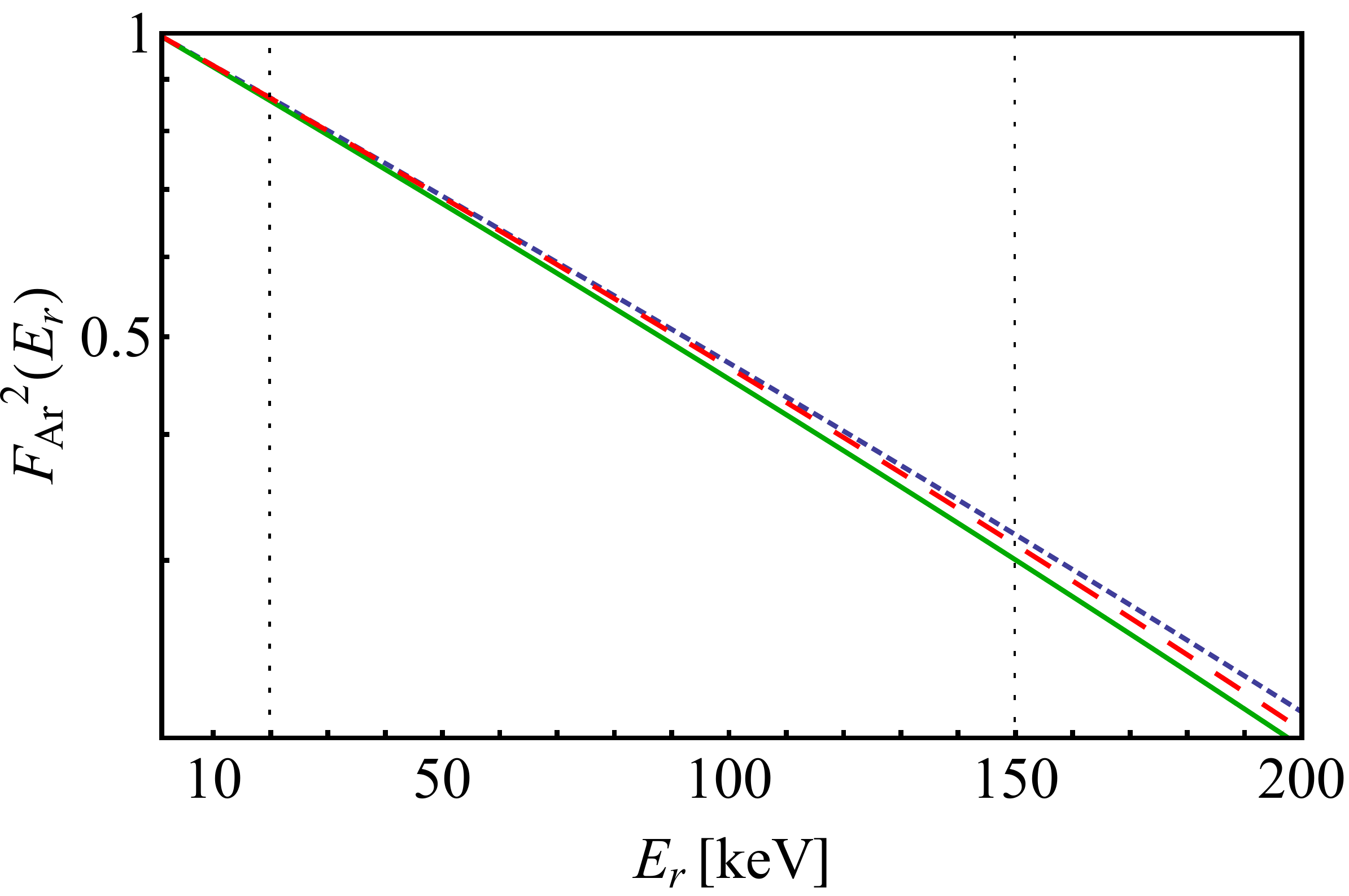}
\includegraphics[height=54mm]{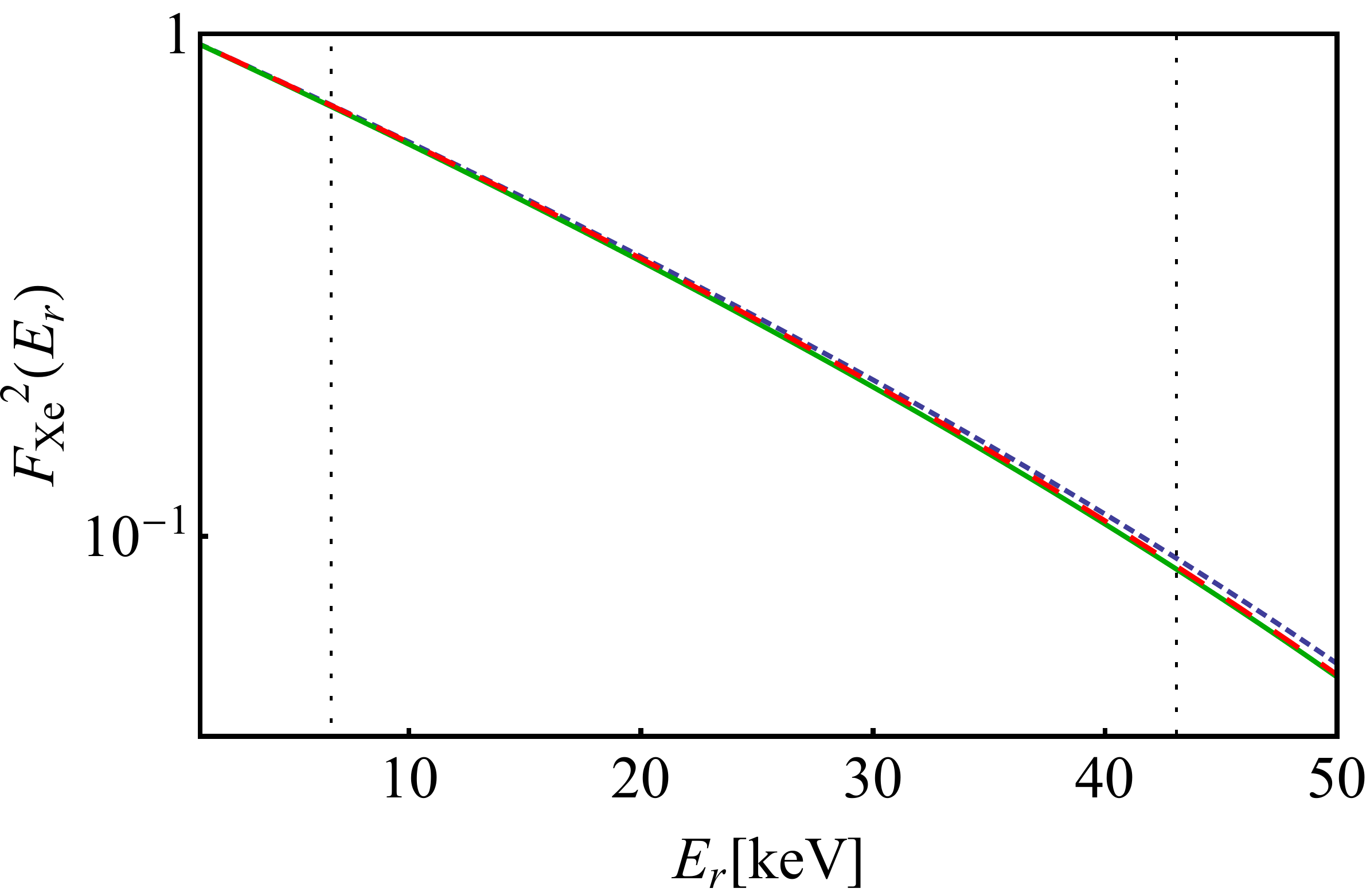}
}
\caption{The Helm (red, dashed), 2PF (blue, fine dashed) and 3PF (green, solid) form factors for \textit{left:} argon-40 \textit{right:} xenon-132 over the energy range relevant to WIMP scattering. The vertical lines show the WIMP search region for each detector.}
\label{figFF}
\end{figure}

Using the parameters given in Table \ref{tableFF}, the 2PF and 3PF form factors for argon and xenon are compared with the Helm form factor in Fig. \ref{figFF}. Given the agreement of the form factors over the relevant WIMP search region of both detectors, we can choose to use the Helm form factor with minimal loss of precision.
Furthermore, it has been shown that small deformations from the assumption of spherical symmetry of the nucleus do not cause any substantial changes to the form factor at low energies \cite{Chen:2011im,Cerdeno:2012ix}.

\begin{table}[ht]
\centering
\caption{Parameters for the charge distributions of argon and xenon}
\begin{tabular}{|l|l|l|}
\hline
 										& 2PF   	 						 & 	3PF		   		 \\
 										\hline
$^{40}\textrm{Ar}$ \cite{Fricke1995177} 		& $c = 3.53 $ fm  			 &	$c = 3.73 \pm 0.05$ fm 		 \\
										& $z = 0.542 $ fm 		 &	$z = 0.62 \pm 0.01$ fm	 \\
										& 			  			 &	$w =-0.19 \pm 0.04$ fm	 \\
\hline				
$^{132}\textrm{Xe} $	\cite{Duda:2006uk,lapikas} &	$c = 3.646 $ fm    	 &	$c = 5.487$ fm   		\\
										& $z = 0.523 $ fm 		 &	$z = 0.557$ fm		\\
										& 				 		 &	$w = 0.219$ fm	 	\\
\hline
\end{tabular}
\label{tableFF}
\end{table}

\bigskip
\subsection{Astrophysical Parameters:  Dark Matter Phase Space Considerations}

The velocity distribution of the WIMPs in the galactic halo is a large source of uncertainty in the calculation of the differential event rate~\cite{Kuhlen2009,Strigari:2009zb,Catena:2011kv,Fairbairn:2012zs,Pato:2012fw}.  Fortunately, while significant uncertainties still remain, there has been significant progress coming from both observational and numerical studies of dark matter in our galaxy.

In considering the impact of the WIMP velocity distribution in the halo on the differential recoil spectrum, we must first transform into the rest frame of the Earth to find the local DM velocity $\mathbf{v}$,
\begin{equation}
\mathbf{v} = \mathbf{v}\,' - \mathbf{v}_e = \mathbf{v}\,' - ( \mathbf{v}_0 + \mathbf{v}_\odot + \mathbf{v}_\oplus),
\end{equation}
where $\mathbf{v}\,'$ is the DM velocity in the Galactic rest frame, and the Earth's velocity $\mathbf{v}_e$ is made up of the galactic rotational velocity, $\mathbf{v}_0$, the Sun's peculiar velocity, $\mathbf{v}_\odot$, and the Earth's orbital velocity about the sun, $\mathbf{v}_\oplus$. The small annual modulation due to $\mathbf{v}_\oplus$  is not considered in this work, and the Sun's peculiar velocity is taken to be $\mathbf{v}_\odot = (10.0,5.23,7.17)$  km/s \cite{Dehnen:1997cq}, where the direction of the three elements of the vector are radially inwards towards the center of the galaxy, in the direction of $v_0$, and upwards from the plane of the galaxy respectively. The choice of $v_0$ is discussed at the end of this subsection.

The standard halo model assumes a singular isothermal sphere of WIMPs, corresponding to a Maxwell-Boltzmann (MB) distribution of velocities,

\begin{equation}
f_{\rm MB}(\mathbf{v}\,') = \frac{1}{v_0^3 \pi^{3/2}}\exp\left[-\frac{\mathbf{v}\,'.\mathbf{v}\,'}{v_0^2}\right].
\label{MB}
\end{equation}
While the singular isothermal sphere is not a good fit to the galactic density profile, the MB velocity distribution actually leads to  somewhat conservative predictions~\cite{McCabe2010}. The advantage of using the MB distribution is that it has an analytical solution to the integral in Eq.~\ref{eqnVminInt}. After converting the integral into an integral of the MB distribution over the speed $v\equiv|\mathbf{v}|$ and the angle between $\mathbf{v}$ and $\mathbf{v}_e$, $\theta$, one finds
\begin{eqnarray}
&\int \frac{f(\mathbf{v})}{|\mathbf{v}|} d^3\mathbf{v}  = \int^{v_{max}}_{v_{min}}\int^{1}_{-1} 2\pi v f(v,\cos\theta)d\cos\theta dv =  \nonumber \\ 
&\frac{1}{2v_e}\left(\mathrm{erf}(\frac{v_e-v_{min}}{v_0})+\mathrm{erf}(\frac{v_e+v_{min}}{v_0})-\mathrm{erf}(\frac{v_e-v_{max}}{v_0})-\mathrm{erf}(\frac{v_e+v_{max}}{v_0})\right),
\end{eqnarray}
where $v_{esc}$ is the galactic escape velocity at the Earth's position. Formally we should truncate the distribution at $v_{esc}$ in the galactic frame before integrating, but setting $v_{max} = v_{esc} + v_e$, the above formula is accurate to a few parts per million.

More realistic velocity distributions can be obtained if one assumes a spherically symmetric spatial distribution and isotropic velocity dispersion of WIMPs in the galactic halo. Specifically, we consider the Hernquist \cite{Hernquist:1990be}, Navarro, Frenk and White (NFW) \cite{NFW}, Burkert \cite{Burkert:1995yz} and Einasto  \cite{Einasto:2009zd,Graham:2005xx} profiles. The NFW profile became the canonical profile for some time, and we include it for direct comparison with the literature. The Einasto profile is similar to the NFW at large radii, but avoids the large central cusp at the Galactic center. The Burkert profile is believed to provide a good description of the DM density profile in dwarf galaxies, and the Hernquist profile has the advantage of an analytic formula for the DM phase-space distribution, as we shall describe shortly. 

In the case of a spherically symmetric velocity dispersion, the velocity distribution can be determined from the gravitational potential according to Eddington's formula~\cite{BT2},
\begin{equation}
F_h(\mathcal{E}) = \frac{1}{\sqrt{8}\pi^2} \left( \int^{\mathcal{E}}_0 \frac{d^2\rho_\chi(r)}{d\Psi^2} \frac{d\Psi}{\sqrt{\mathcal{E}-\Psi}} + \frac{1}{\sqrt{\mathcal{E}}} \left(\frac{d\rho_\chi(r)}{d\Psi}\right)_{\Psi=0} \right).
\label{eddington}
\end{equation}
Here the relative potential $\Psi(r)$ and the relative energy $\mathcal{E}$ are defined as
\begin{eqnarray}
&\Psi(r) = - \Phi(r) \,\,\,\mathrm{and} \\
&\mathcal{E} = -E = \Psi(r)-E_k, \\ \nonumber
	\label{potential}
\end{eqnarray}
where $\Phi$ is the gravitational potential, and $E$ and $E_k$ are the total and kinetic energy respectively. 

The velocity distribution determined from Eq.~(\ref{eddington}) is 
self-consistent, and more likely to describe the behavior of the DM particles
in the Milky Way than the MB shape in Eq.~(\ref{MB}). However, a few words
of caution are in order. Our DM halo is assumed to be self-gravitating, 
i.e. we find the gravitational potential $\Phi$ solving Poisson's equation
for the particular DM density profile under consideration. In doing so, we
are disregarding the effect of baryons, which deepen the gravitational well
and affect the evolution of the DM density through dissipative processes.
Disregarding the latter, the additional gravitational pull due to the
baryons can be included
by using spherical approximations to the baryonic bulge and 
disk~\cite{Strigari:2009zb}. Alternatively, one can resort to hydrodynamic 
numerical simulations, which show that dissipational baryonic processes can 
increase the local DM density and broaden the velocity 
distribution~\cite{Kuhlen:2013tra}, although the net effect on the
time-averaged scattering rate is only midly changed. 
These effects are certainly important,
specially when comparing the results of different experiments, and we
plan to include them in a future work. 
Nonetheless, for our present purposes, the range of different
shapes for the velocity distribution and the uncertainty that we allow for
$\rho_\chi$ are sufficient to capture the influence of baryons.

Seeking to determine the local dark matter density, Catena and Ullio  \cite{Catena2011} used a Bayesian approach to constrain the 7 (8 for Einasto) parameters needed to model the Milky Way. These parameters are as follows: our distance from the center of the galaxy; two dark matter halo parameters (the virial mass and a dimensionless virial scale, plus a halo profile shape parameter for Einasto); three baryonic parameters; and a parameter to encode the anisotropy of halo stars (see Ref.~\cite{Catena2011} for definitions). 
The analytic phase-space distribution for the Hernquist profile can be  obtained using Eq.~\ref{eddington} in combination with the density profile and potential,
\begin{align}
\rho_H(r) &= \frac{M_\mathrm{MW} a}{2 \pi r(r+a)^3},\\
a &= \frac{\sqrt{G_N\,M_{\mathrm{MW}} R_0} - R_0 v_0}{v_0},\\
\Phi &= - \frac{G_N\,M_{\rm{MW}}}{r+a},
\end{align}
where $R_0$ is the distance from the Sun to the center of the Galaxy and $M_{MW}$ is the mass of the Milky Way,
giving \cite{Hernquist:1990be}
\begin{align}
f(q)&=\frac{\left(8 q^4-8 q^2-3\right) q \sqrt{1-q^2} \left(1-2 q^2\right)+3 \sin^{-1}(q)}{\left(1-q^2\right)^{5/2}},\\
q&=\sqrt{\frac{a \epsilon }{G_N \,M_{\rm MW}}},\\
\epsilon&=\frac{G_N \,M_{\rm MW}}{a+R_0}-\frac{1}{2} \left(\mathbf{v}\,'.\mathbf{v}\,'\right).
\end{align}
We adopt a value of $v_0/R_0 = 29.45$ km/s/kpc \cite{Reid:2004rd}. $M_{MW}$ is determined from $\rho_\chi$ and $v_0$ following the technique in Hernquist (1990) \cite{Hernquist:1990be}.
 Finally, while the effect of microhalos on direct detection experiments has been shown to be minimal \cite{Schneider:2010jr}, N-body simulations of galactic halos do show a departure on small scales from the standard smooth isothermal model. Thus, we also consider here the results of the Via Lactea numerical simulation \cite{Kuhlen2009}, for comparison with the analytic model estimates. 
 
\begin{figure}[pth]
\centering
\mbox{
\hspace{-9mm}
\includegraphics[height=38mm]{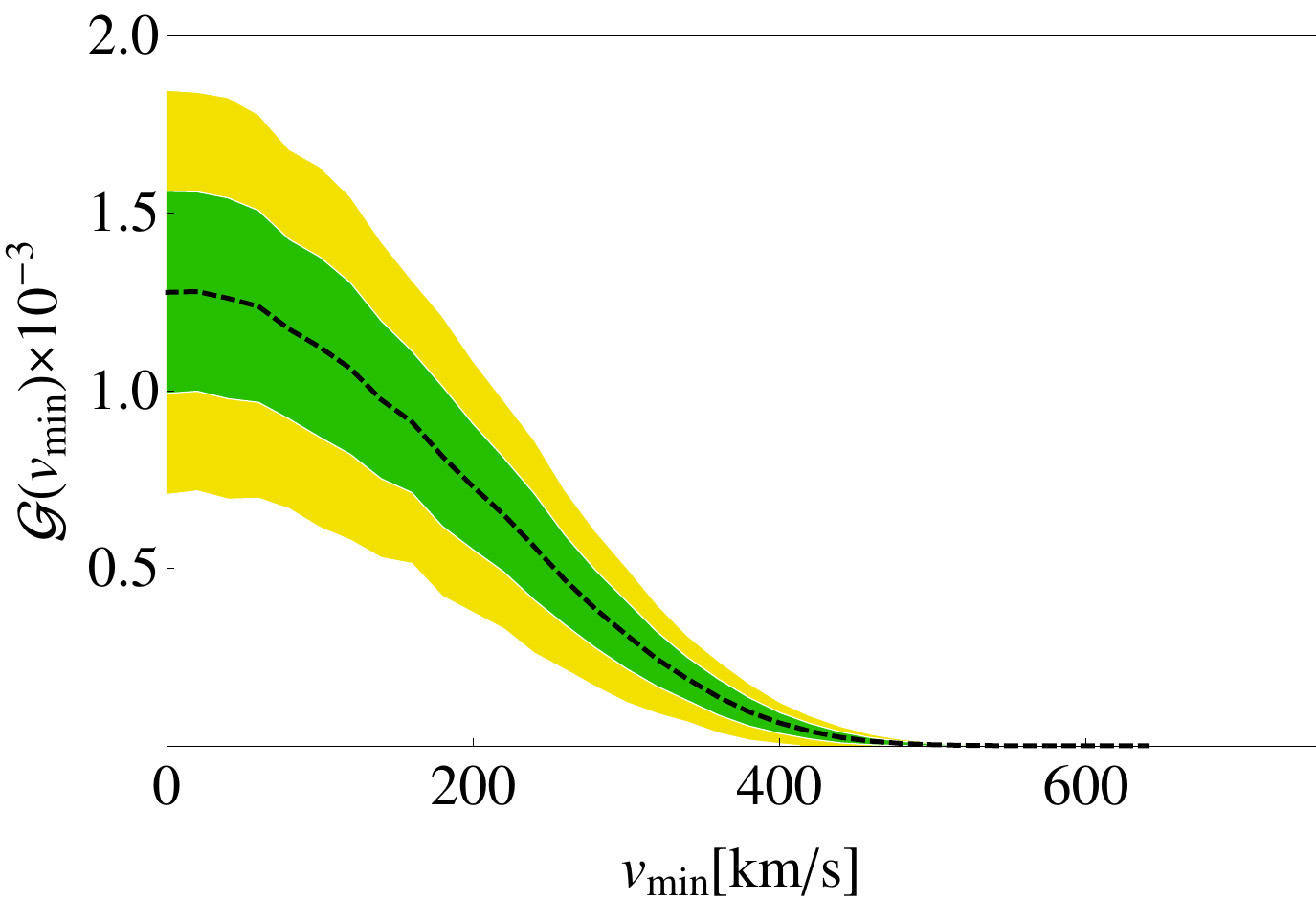}
\includegraphics[height=38mm]{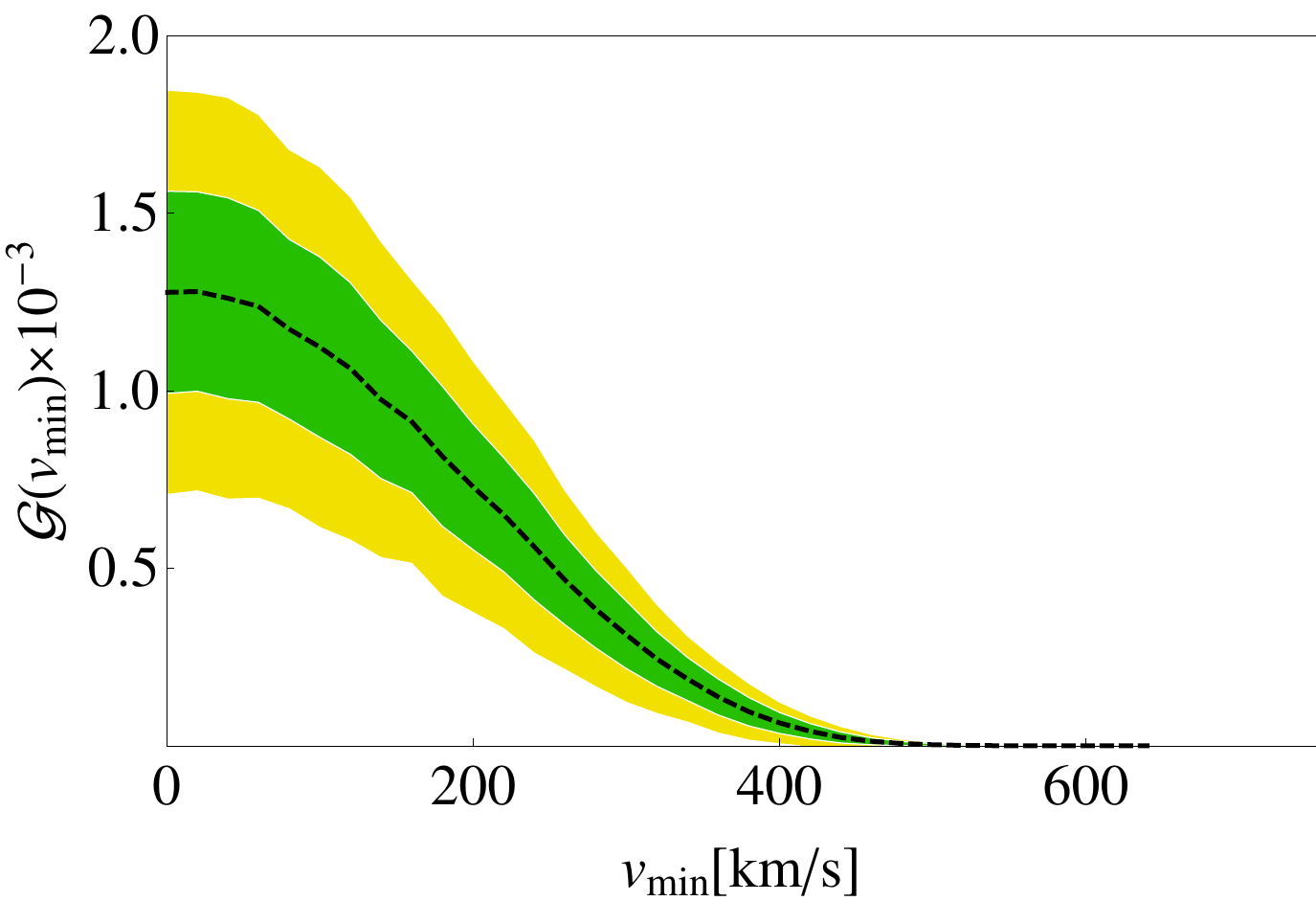}
\includegraphics[height=38mm]{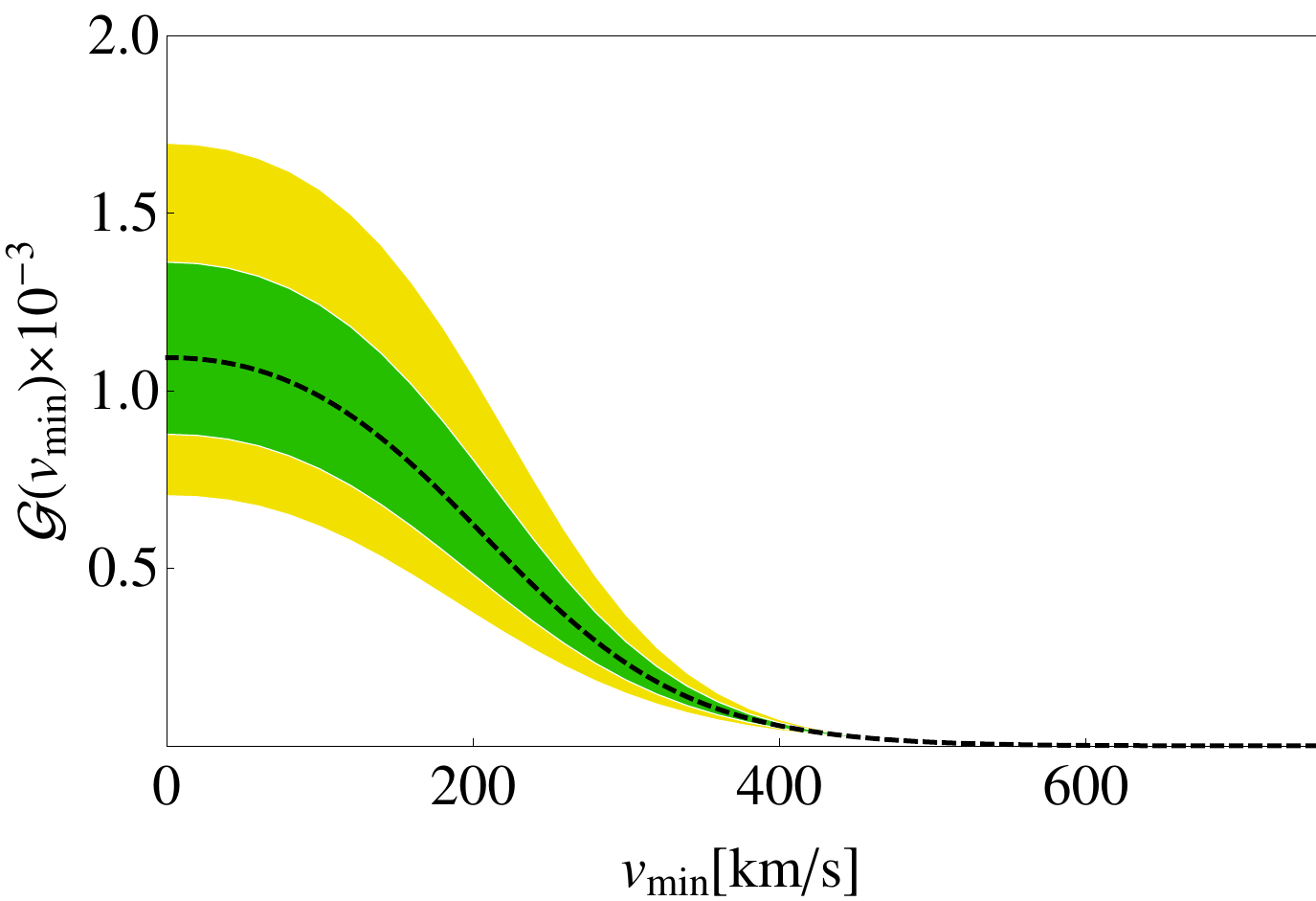}
}
\caption{Numerical results of Eq.~\ref{eqnVminInt} for, from left: MB, Hernquist, and Via Lactea profiles. The black dashed curve shows the mean value, while the green and yellow regions show one- and two-sigma errors. Note the errors here are larger than in Fig.~\ref{figFvNum2} since only $v_0$, $\rho_\chi$, and $v_{esc}$ (and for Hernquist, $R_0$, distance to the center of the galaxy) are used to constrain these models.}
\label{figFvNum1}
\end{figure}

\begin{figure}[htp]
\centering
\mbox{
\hspace{-9mm}
\includegraphics[height=38mm]{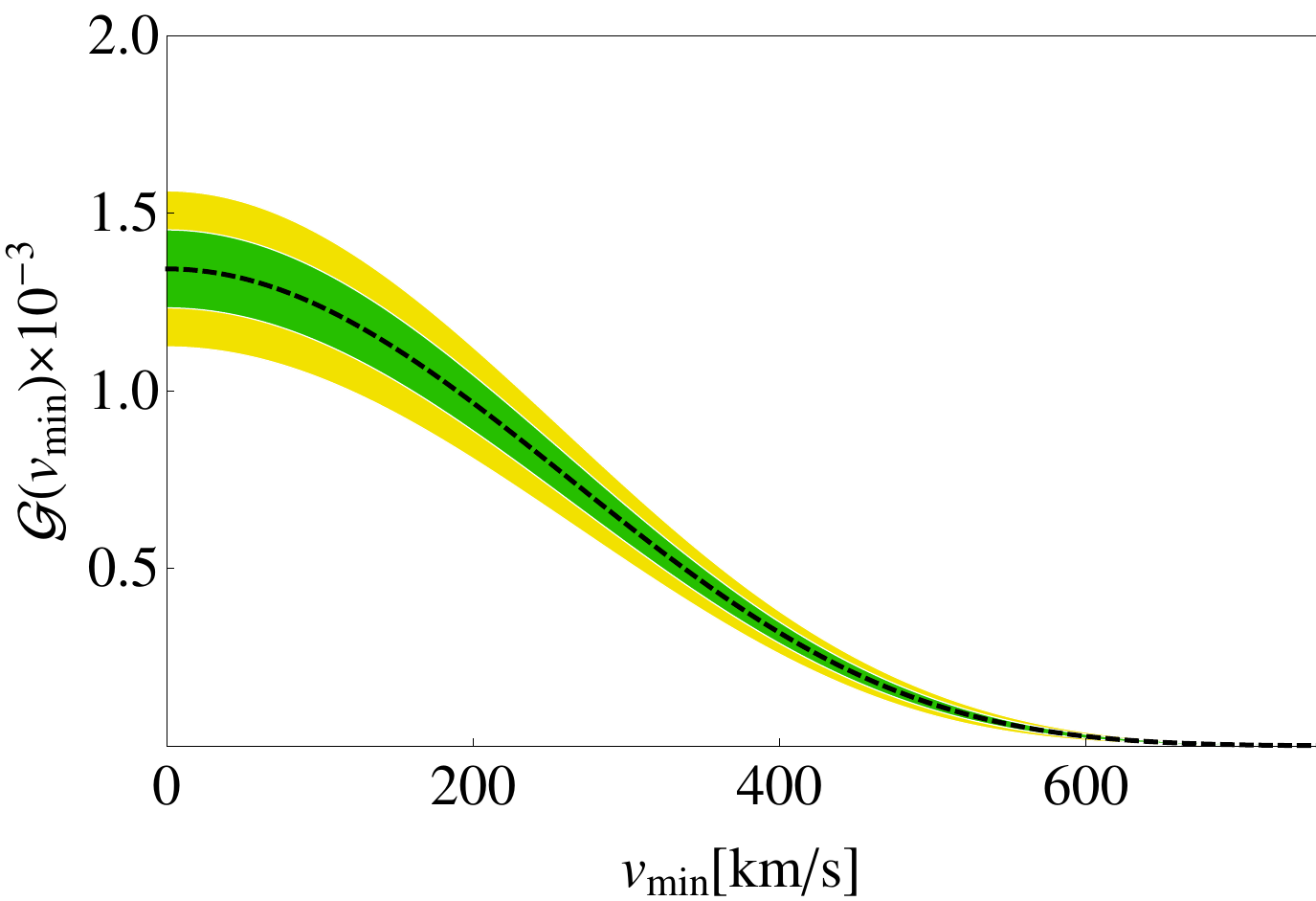}
\includegraphics[height=38mm]{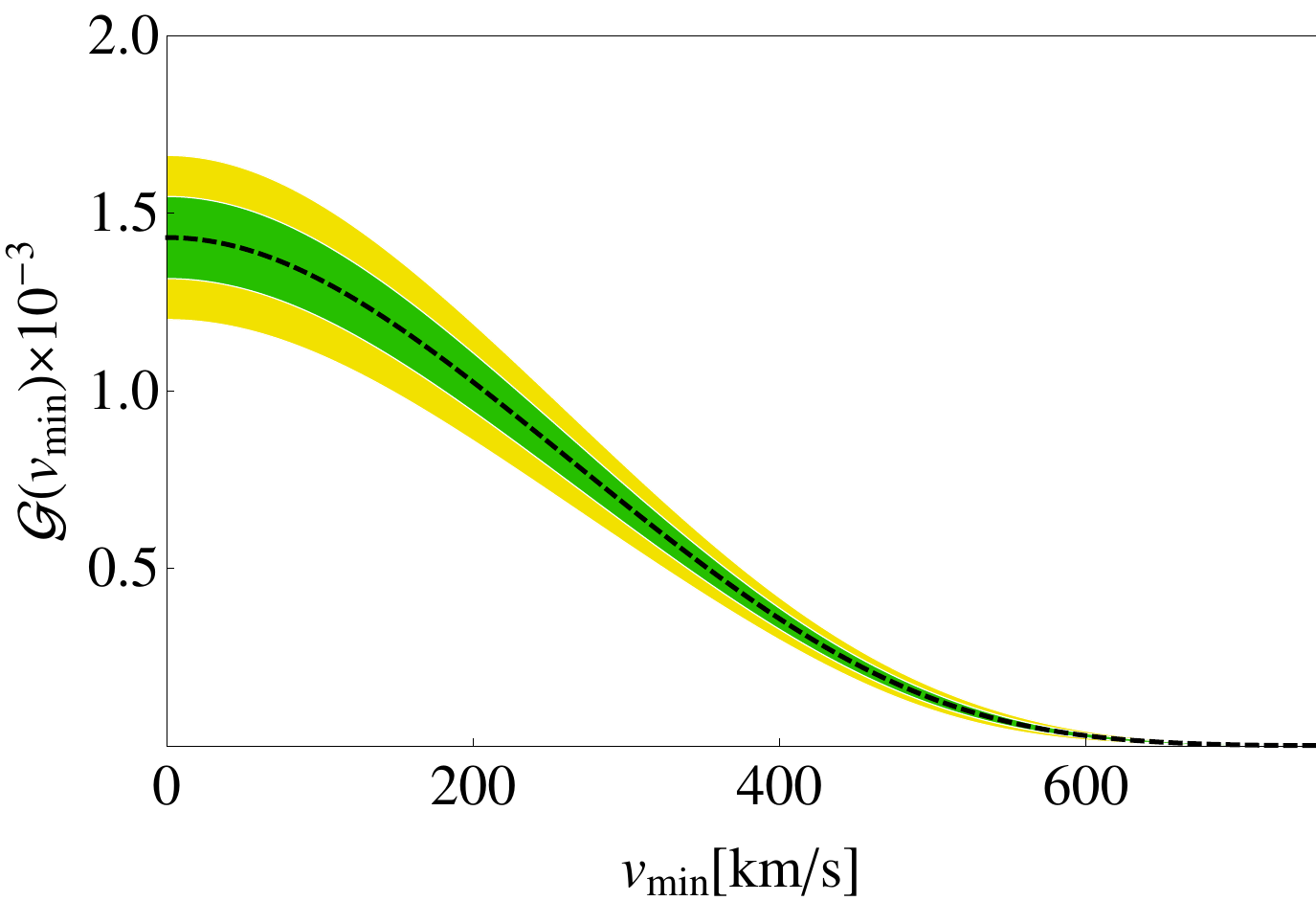}
\includegraphics[height=38mm]{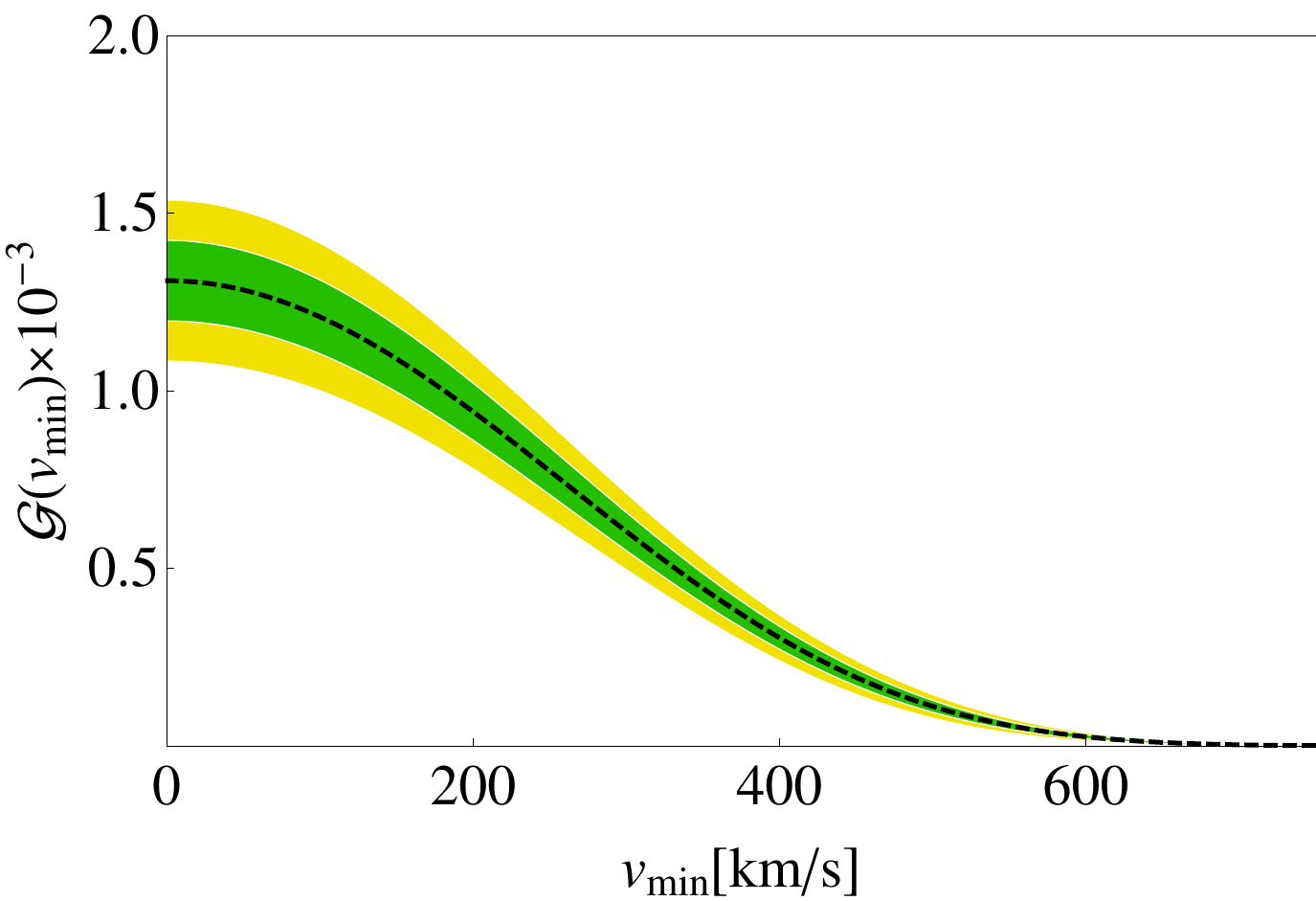}
}
\caption{Numerical results of Eq.~\ref{eqnVminInt} for each of the velocity distributions from \cite{Catena:2011kv}. From left: NFW, Burkert and Einasto profiles. The black dashed curve shows the mean value, while the yellow and green regions show one- and two-sigma errors.}
\label{figFvNum2}
\end{figure}

 For an illustrative comparison of how uncertainties in these distributions affect the WIMP scattering rate, each of the distributions is integrated by Eq.~\ref{eqnVminInt} and the results are shown in Fig.~\ref{figFvNum1} and Fig.~\ref{figFvNum2}. The MB, Herquist and Via Lactea distributions use the standard astrophysical assumptions of $v_0 = 220\pm20$ km s$^{-1}$, $v_{esc} = 544^{+ 64}_{-46}$ km/s$^{1}$ and $\rho_\chi = 0.3\pm0.1$ GeV/cm$^{2}$ \cite{Green:2011bv}.  Note that  there is considerable variation in the favoured values of $v_0$ and $\rho_\chi$ (see \cite{McMillan:2009yr,Salucci:2010qr,Green:2011bv}). The large uncertainties we adopt cover most of the proposed range of these parameters.  Also note that other halo models are designed to more accurately model the details of the data and thus have a smaller range of quoted uncertainties in the phase space distribution.

\subsection{Backgrounds} 
Ultimately, it is the background rate that sets the lower limit of observable signal rates, so that significant attention must be paid to both shielding the detector from unwanted radioactive backgrounds, and also to devising methods to distinguish between possible signal and background events, in particular to distinguish candidate WIMP events  which involve single scatter nuclear recoils from multiple scatter nuclear events and electronic recoils. 

The XENON100 detector was able to achieve a pre-discrimination background rate of $5.3 \times 10^{-3}$ dru (events/kg/day/keV$_{\rm n.r.}$) \cite{Aprile:2012nq}. For the future xenon component of the DARWIN detector and argon DarkSide-50 detector, the pre-discrimination electronic background goal is $10^{-6}$ dru (not including the solar neutrino background) and $\mathcal{O}(1)$ dru respectively. In both cases  the background is assumed to be constant in energy, and the radioactive nuclear recoil background is subdominant. 

Liquid scintillators discriminate nuclear and electronic recoils via prompt vs.~delayed signal cuts and/or pulse shape analysis. While the electronic recoil background in argon detectors is currently much larger, electronic recoils in argon can be discriminated at a rate of 1 part in 10$^7$ \cite{Benetti:2007cd}, compared with 2.5 parts in 10$^3$ for xenon. To provide a coincident detection and maximize complementarity, the argon detector must be as sensitive as the xenon detector, requiring a factor of 100 reduction in the argon background, which could be achieved through the use of low radioactivity argon \cite{Back:2012pg}.

\begin{figure}[ht]
\centering
\includegraphics[width=80mm]{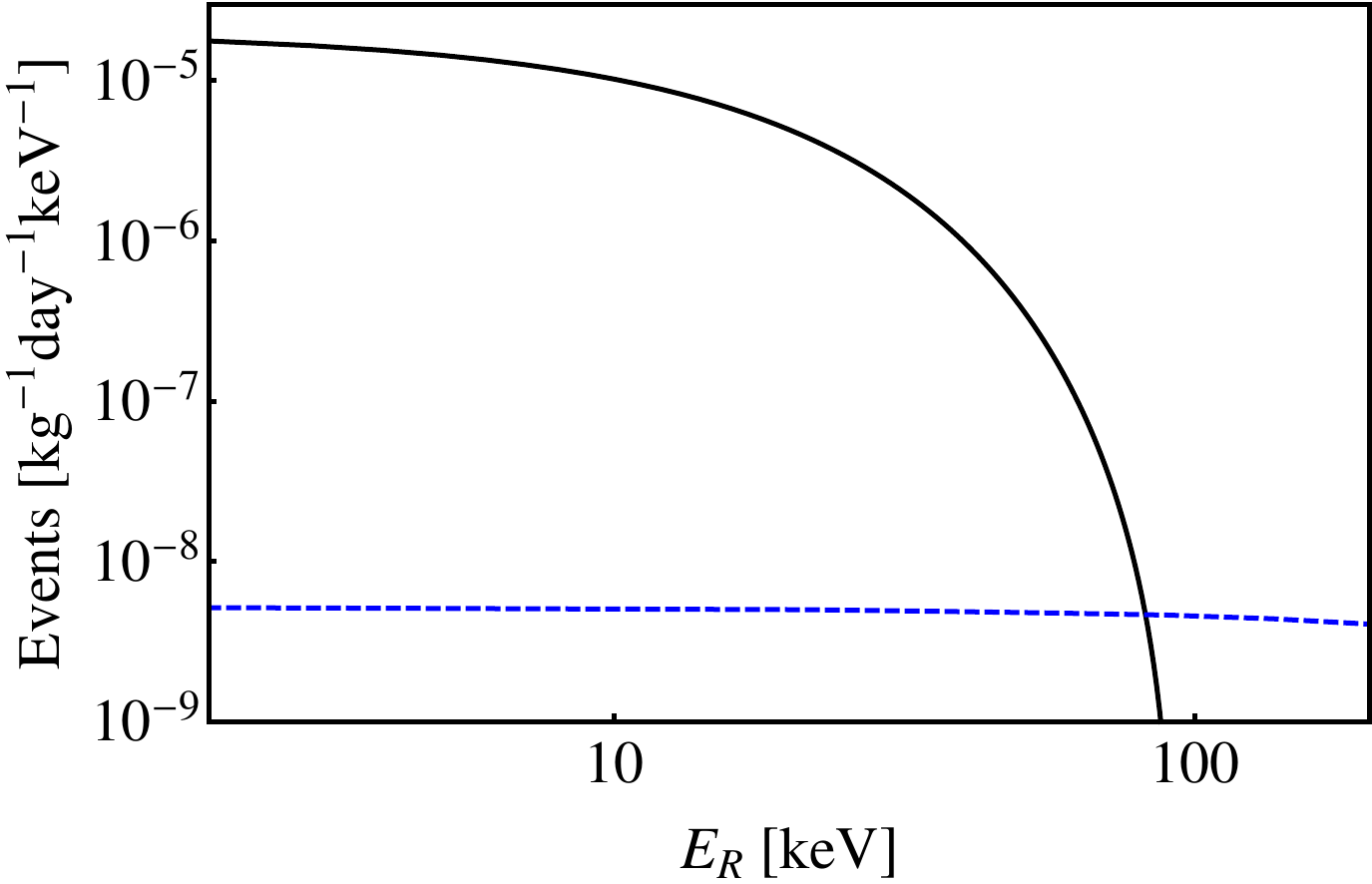}
\caption{The differential event rate in a xenon detector for a 100GeV WIMP with $\sigma_{\chi p}=3\times10^{-46}$~cm$^2$ using standard astrophysical assumptions (black solid) and the irreducible (after 99.75\% rejection) neutrino backgrounds (blue dashed) \cite{Baudis:2012bc} }
\label{figSpectrum}
\end{figure}

Beyond intrinsic detector backgrounds, there is one ultimate background that is irremovable, and puts a lower limit on the scattering cross section sensitivity of WIMP dark matter detection experiments of the type considered here.  This is the solar neutrino background, which comes in at a level of $\sigma = 10^{-48} \mathrm{cm}^2$.
In particular, elastic scattering of solar pp-neutrinos from electrons provides a flat background which cannot be feasibly screened. While electronic recoils can be discriminated and rejected, at some level, below that level the remaining spectrum (see Fig.~\ref{figSpectrum}) is irreducible. This corresponds to a rate of $1.8\times10^{-4}$ events/tonne/day in the xenon WIMP search region~\cite{Baudis:2012bc}. To obtain a rate for argon detectors one must scale the xenon spectrum by $\frac{Z_{Ar}A_{Xe}}{Z_{Xe}A_{Ar}} = 1.096$.  Due to the considerations described in the previous paragraph, however, this will be a sub-dominant component of the background in argon. 

\section{Projected Sensitivity}

\subsection{Projected Experimental Upper Limits}
To estimate the sensitivity of future experiments we construct 90$\%$ exclusion limits using the profile likelihood method on a representative `Asimov' dataset~\cite{Aprile:2011hx,Cowan:2010js}. This method utilizes the test statistic,
$$
q_\sigma =
\begin{cases}
   -2 \mathrm{log} ( \lambda(\sigma) ) & \sigma  \geq \hat\sigma \\
   0	& \sigma < \hat\sigma \\
\end{cases}
$$
where $\lambda$ is the profile likelihood ratio,
\begin{equation}
\lambda(\sigma) = \frac{\mathcal{L}(\sigma,\hat{\hat{\theta}})}{\mathcal{L}(\hat{\sigma},\hat{\theta})}.
\end{equation}
Here $\theta$ represents all of the uncertain parameters that enter the likelihood, $\hat\sigma$ and $\hat\theta$ denote that the likelihood has been maximized with respect to those parameters and $\hat{\hat\theta}$ denotes the likelihood has been maximized for the given $\sigma$. The likelihood function is a product of the probabilities of having observed $A_i$ events, given the expected $E_i$ events, for a given energy bin,
\begin{equation}
\mathcal{L}(\sigma,\theta) = \prod^N_{i=1} P(E_i(\sigma,\theta),A_i).
\label{eqnLikelihood}
\end{equation}
 The expected number of events $E_i$ is the sum of both the expected DM recoil events and the background events in that energy bin. We define the WIMP search regions to be 6.6-43 keV for xenon and 20-150 keV for argon; the regions are split up into bins of width 5 keV (in lieu of smearing). The detector parameters are summarized in Table \ref{tableDetectors}.
\begin{table}[t]
\caption{These detector parameters are motivated by current experiments and expected performance of future detectors \cite{Benetti:2007cd,DarkSide,Aprile:2012zx}.  The backgrounds are assumed to be constant in energy. }
\begin{tabular}{|l|r|r|}
\hline
                                       & Xenon             &   Argon \\
\hline
Nuclear recoil acceptance              & $40\%$            & $50\%$ at 35keV, $100\%$ \textgreater60 keV \\
Total background (post-discrimination) & $6\times10^{-9} $ dru & $2.3\times10^{-9}$ dru\\
WIMP search region                     & 6.6-43 keV        & 20-150 keV         \\
\hline

\end{tabular}
\label{tableDetectors}
\end{table}

\begin{figure}[pth]
\centering
\mbox{\hspace{-10mm}
\includegraphics[width=60mm]{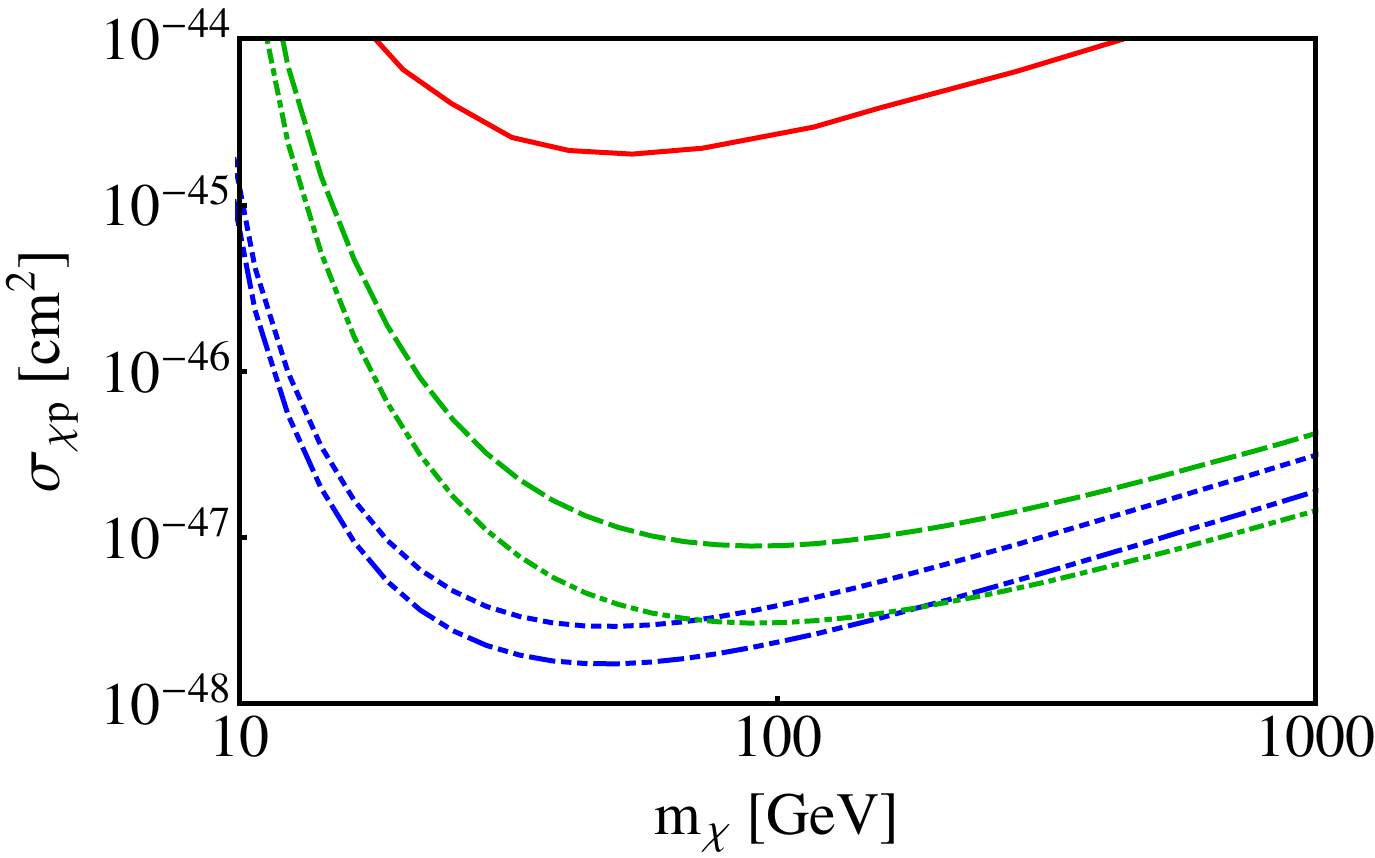}
\includegraphics[width=60mm]{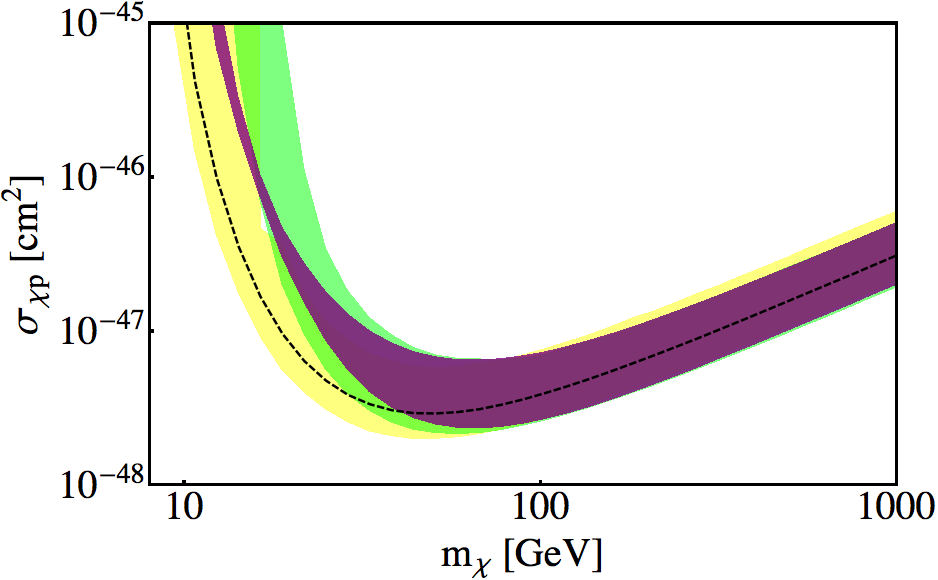}
\includegraphics[width=60mm]{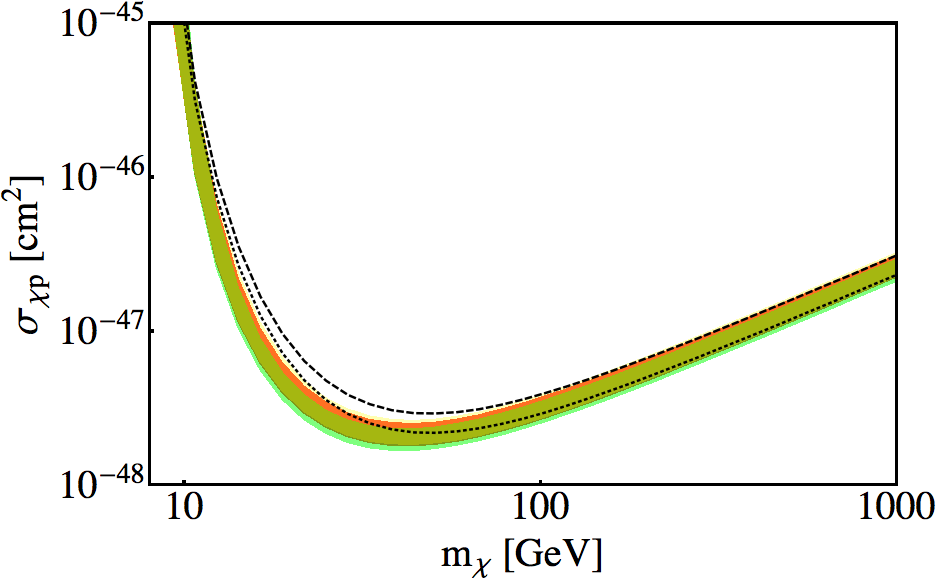}
}
\caption{\textit{Left:} comparison of exclusion limits for a 10 (blue, dotted) and 20 (blue, dot-dot-dashed) tonne-years xenon exposure, 20 (green, dashed) and 30 (green, dot-dashed) tonne-years argon exposure, and the current best limits set by XENON100 (red, solid) \cite{Aprile:2012nq} (standard astrophysical assumptions). \textit{Center:} the effect of astrophysical uncertainties on a 10 tonne-years xenon exposure with neutrino-only backgrounds for different WIMP halo profiles compared to MB with standard assumptions (black dashed):  MB (yellow), Herquist (green), Via Lactea II (purple) and \textit{right:} Einasto (yellow), NFW (red), Burkert (green) and MB with $\rho_\chi=0.4$ GeV cm$^{-3}$ (dotted).}
\label{figNeutrinoBGlimits}
\end{figure}

 Typically, the XENON collaboration exposes their detector for the length of time expected to produce a single background event~\cite{Aprile:2012nq}. With this in mind, the solar neutrino background limits exposure to around 10 tonne-years in xenon. The limits obtained for several exposures of xenon and argon compared to the final XENON100 limits are shown in Fig.~\ref{figNeutrinoBGlimits} (left). Note that to achieve comparable sensitivity, a larger fiducial volume of argon is necessary compared with xenon. Unless the neutrino background can be unambiguously subtracted or otherwise discriminated (e.g.~via the use of directional information as described in \cite{Copi:2002hm}), these limits approximately represent the floor to the sensitivity of the current xenon liquid scintillator design.  Fig.~\ref{figNeutrinoBGlimits} (center and right) shows the effect of the uncertainty of the phase-space density on a 10 tonne-years xenon exposure. The NFW, Einasto and Burkert profiles enforce more stringent limits because they favor a local WIMP density of $\rho_\chi=0.4$ GeV cm$^{-3}$~\cite{Catena:2011kv}. Thus the standard MB assumptions are conservative in comparison to these more realistic profiles.
	
\subsection{Signal Simulation and Parameter Reconstruction}
After specifing a WIMP model:
\begin{itemize}
\item WIMP mass $m_\chi$,
\item proton cross section $\sigma_{\chi p}$,
\item isospin violating factor $\frac{f_n}{f_p}$, and
\item inelastic parameter $\delta$,
\end{itemize}
we generate an Asimov dataset of recoil events according to the differential event rate Eq.~\ref{eqndN}. The simulated events are binned as defined in the previous section and the MultiNest sampler~\cite{multinest08} is used to reconstruct the WIMP model parameters (or a subset therein). MultiNest returns the full posterior probability distribution via Bayes theorem,
\begin{equation}
\mathcal{P}(\theta,D|I) = \frac{\mathcal{L}(D|\theta,I)\pi(\theta,I)}{\epsilon(D,I)},
\end{equation}
where the likelihood function is as previously defined in Eq.~\ref{eqnLikelihood}, and $\pi$ and $\epsilon$ are the prior probabilities and Bayesian evidence respectively. The types of priors used are given in Table \ref{tablePriors}. We then marginalize the posterior probability over all parameters except the WIMP mass and proton cross section. Except where otherwise noted, the inelastic and isospin violating parameters were fixed to $\delta=0$ keV and $\frac{f_n}{f_p}=1$ and not allowed to vary in the reconstruction.

\begin{table}[ht]
\centering
\caption{The chosen priors for the WIMP sampling parameters and the standard astrophysical parameters, motivated by \cite{Koposov:2010, Siebert:2012ka, Weber:2009pt}, errors denote 1-sigma intervals.}
\begin{tabular}{|c|c|c|}
\hline
Parameter & Range & Prior \\
\hline
$m_\chi$			&  $1-2000$ GeV 	              	& 	log  \\
$\sigma_{\chi p}$	&  $10^{-48}-10^{-42}$ cm$^2$	&	log  \\
$\frac{f_n}{f_p}$  &  -$4-4$			 		&  linear  \\
$\delta$			&  $0-100$ keV		 		&  linear   \\
\hline
$v_0$			& $220\pm20 $ km/s      		& Gaussian  \\
$v_{esc}$	   	& $544\pm40 $ km/s	          & Gaussian  \\
$\rho_\chi$		& $0.3\pm0.1$ GeV/cm$^2$	& Gaussian \\
\hline
\end{tabular}
\label{tablePriors}
\end{table}

\begin{figure}[tp]
\centering
\mbox{
\includegraphics[width=80mm]{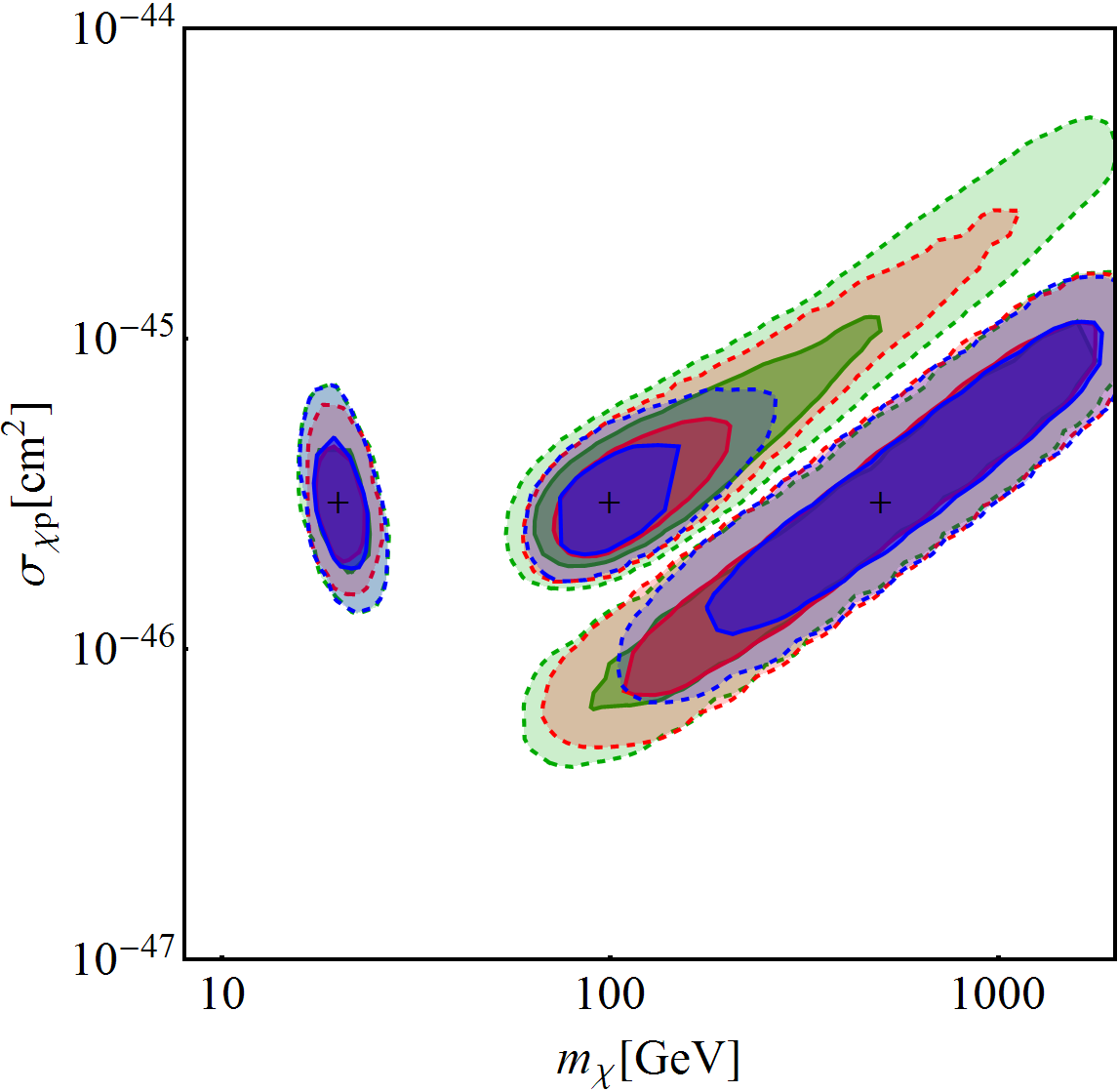}
\includegraphics[width=80mm]{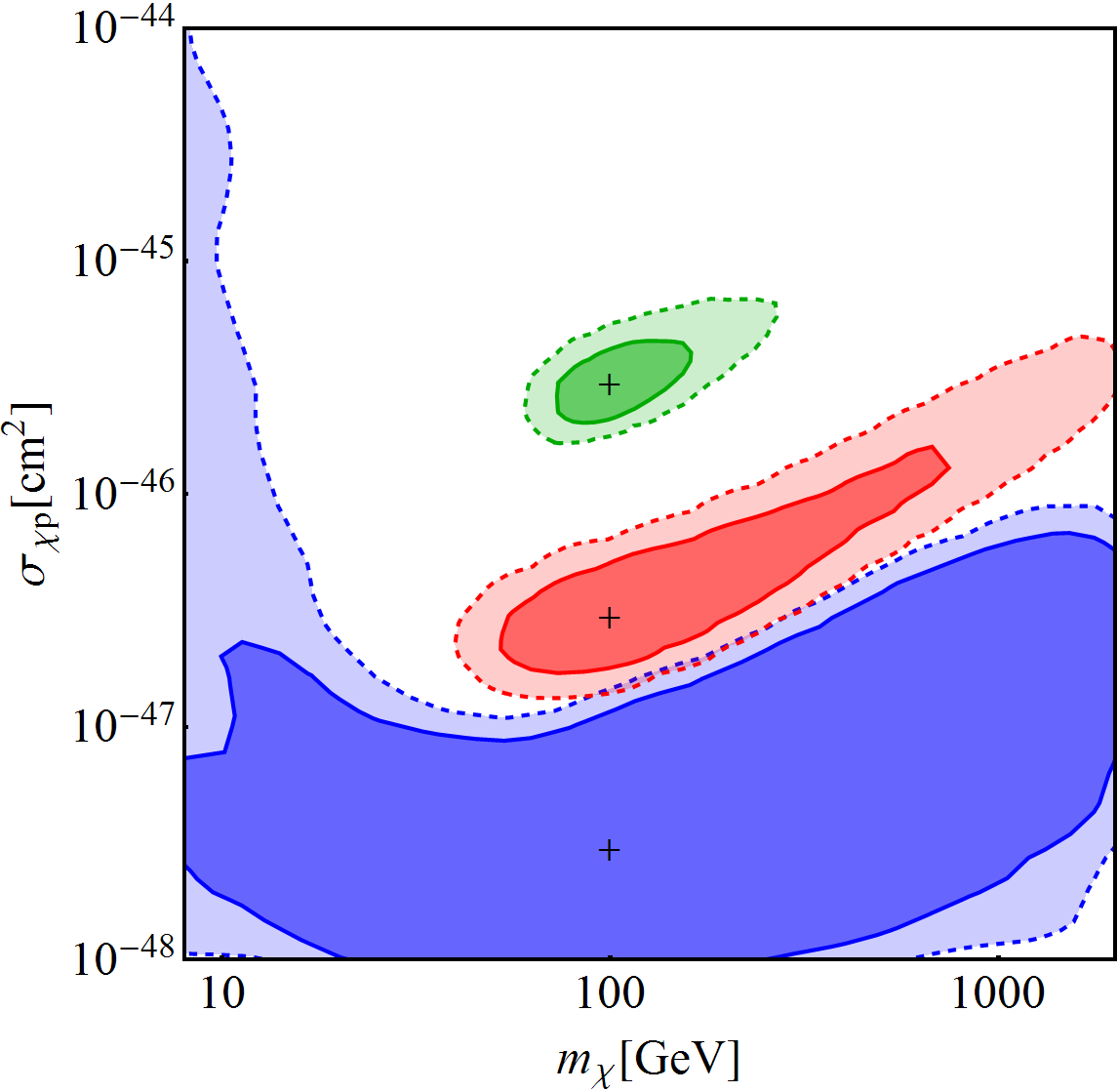}
}
\caption{One- and two-sigma credible regions of the marginal posterior probabilities for simulations of WIMPs with \textit{left:} $\sigma_{\chi p}=3\times10^{-46}$ cm$^2$, and masses 20GeV, 100GeV and 500GeV, for exposures of 10 tonne-years xenon (green), 20 tonne-years xenon (red) and 10 tonne-years xenon plus 20 tonne-years argon (blue). \textit{Right:} $\sigma_{\chi p}=3\times10^{-46}$ cm$^2$ (green),  $\sigma_{\chi p}=3\times10^{-47}$ cm$^2$ (red) and  $\sigma_{\chi p}=3\times10^{-48}$ cm$^2$ (blue) for an exposure of 10 tonne-years xenon plus 20 tonne-years argon. The `+' indicates the simulated model.}
\label{figReconstruct2}
\end{figure}

To test the complementarity of a xenon and argon detector, WIMP events with $\sigma_{\chi p}=3\times10^{-46}$ cm$^2$ and masses of 20, 100 and 500 GeV were simulated for xenon and xenon plus argon detector configurations. The resulting detector reconstructions are shown in Fig.~\ref{figReconstruct2} (left). The Helm form factor and MB distribution were used (with uncertainties marginalized). The results show, with the detectors working together, that complementarity does provide a small improvement across the whole mass range, but most significantly at 100GeV (approximately the crossover between the different detector sensitivities). 
It is interesting to contrast this with the improvement gained by increasing the exposure of the xenon detector alone, either through increasing the exposure time or fiducial volume also shown in Fig.~\ref{figReconstruct2} left. This allows us to compare the increase in sensitivity due to the complementarity between the targets versus the improvement due to the increased exposure.
We can see that by using the two detectors there is an improvement in the $2\sigma$ error in the reconstructed mass, but at $1\sigma$ the improvement is very minor. Note that where degeneracies exist (e.g.~the $m_\chi=500$ GeV reconstruction in Fig.~\ref{figReconstruct2}) or the statistics are low (e.g.~the $\sigma_{\chi p}=3\times10^{-48}$ cm$^2$ reconstruction of Fig.~\ref{figReconstruct2} right) the apparent cutoff of the credible regions at the edges of the graphs are artifacts of our mass and cross section priors ($M \le 2$ TeV, $\sigma \ge 10^{-48}$ cm$^2$).


\begin{figure}[htp]
\centering
\mbox{
\includegraphics[width=80mm]{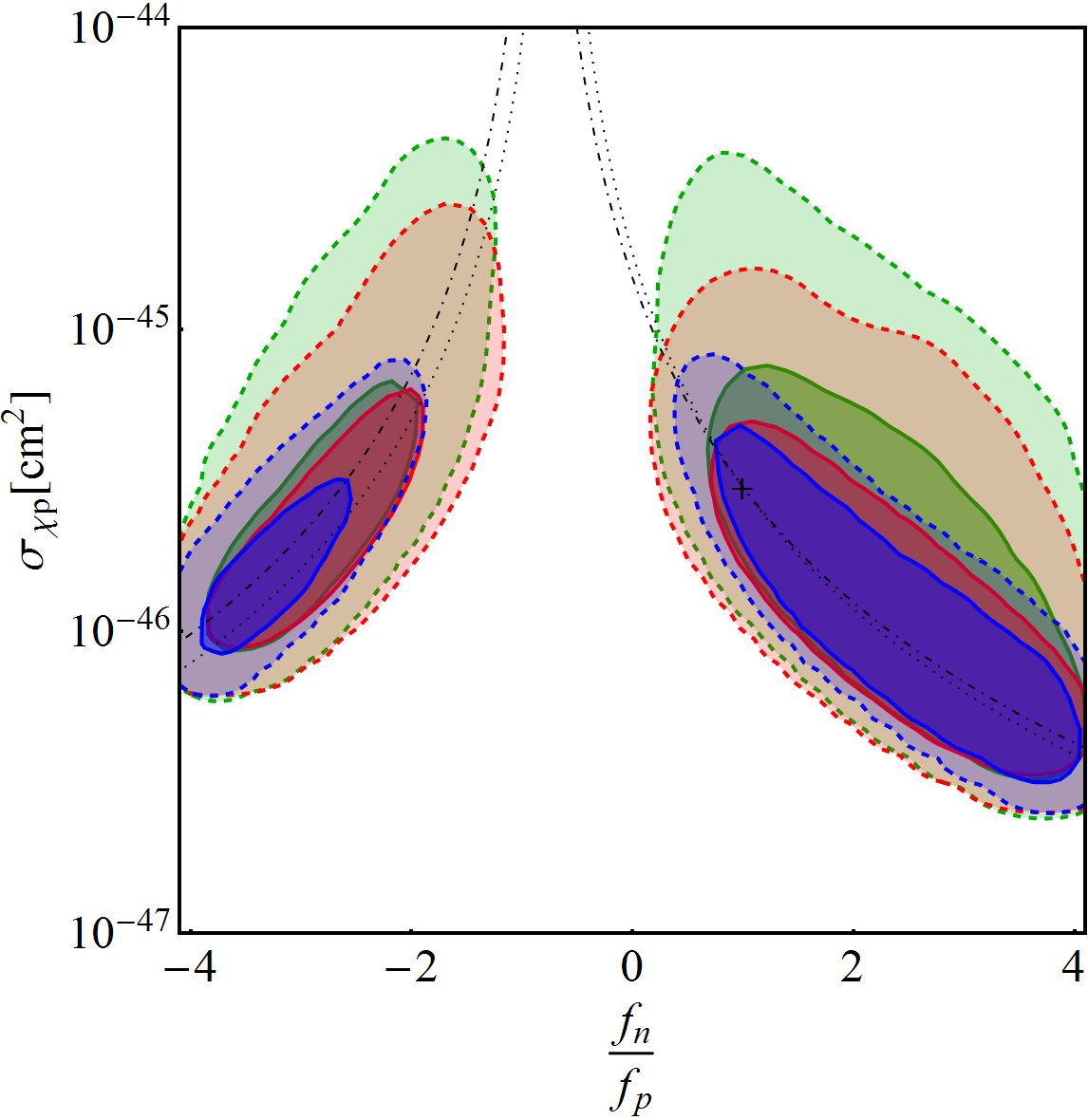}
\includegraphics[width=80mm]{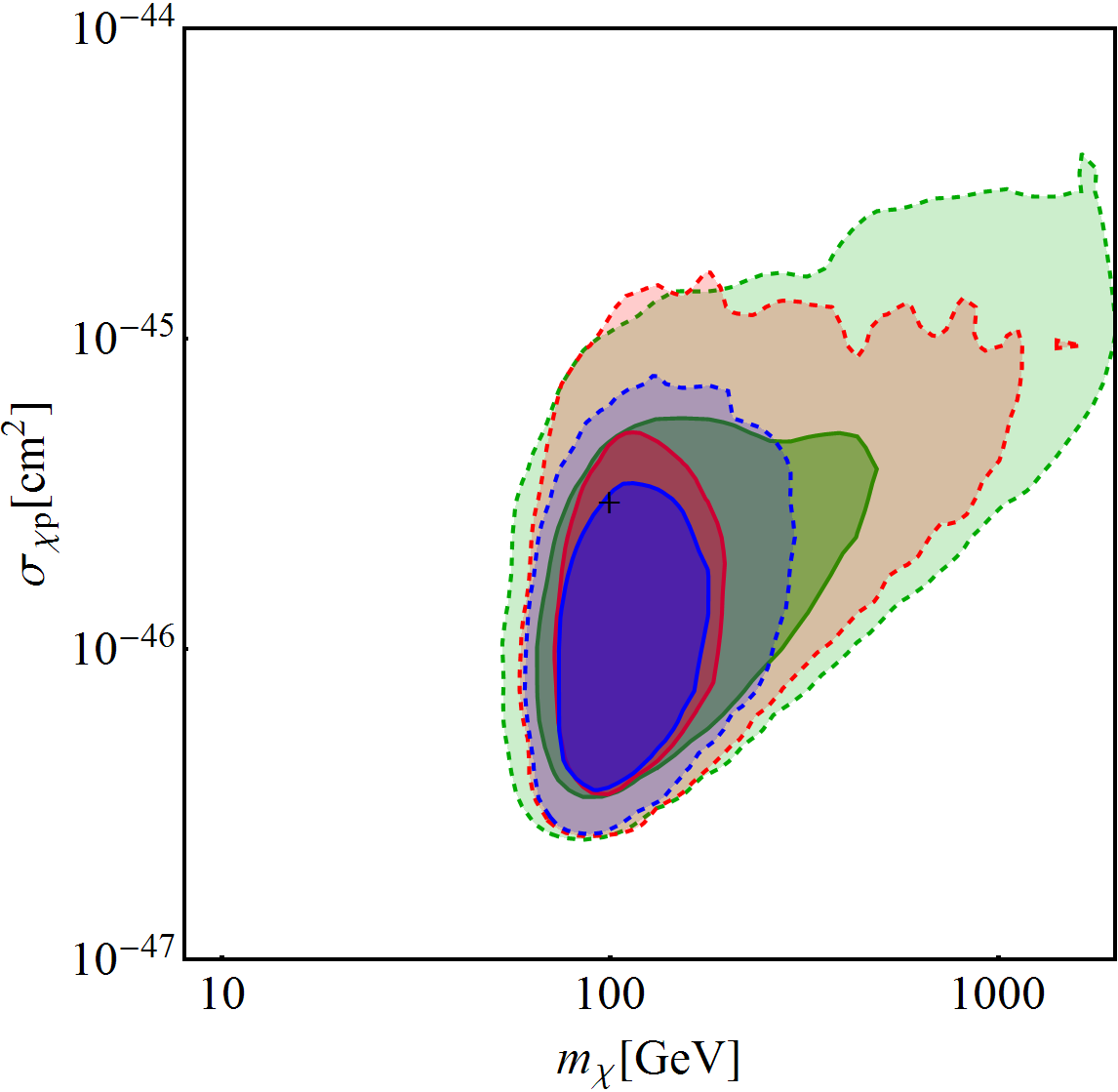}
} 
\caption{One- and two-sigma credible regions of the marginal posterior probabilities for simulations of WIMPs with $\sigma_{\chi p}=3\times10^{-46}$ cm$^2$, $m_\chi=100$ GeV, $\frac{f_n}{f_p}=1$ and $\delta = 0$ keV. 
In both figures the isospin violating parameter $\frac{f_n}{f_p}$ is allowed to vary during reconstruction. 
The dotted and dot-dashed curves show the degeneracy between $\sigma_{\chi p}$ and $\frac{f_n}{f_p}$ for argon and xenon respectively. 
Shown are exposures of 10 tonne-years xenon (green), 20 tonne-years xenon (red) and 10 tonne-years xenon plus 20 tonne-years argon (blue). 
(\textit{Left}): Reconstruction in the $\sigma_{\chi p} - \frac{f_n}{f_p}$ plane. 
(\textit{Right}): Reconstruction in the $\sigma_{\chi p} - m_\chi$ plane (note that the spikes are due to sampling error in the reconstruction).}
\label{figFnFp}
\end{figure}

In the case of isospin violating interactions, we still simulate a WIMP with $\frac{f_n}{f_p}=1$, but now allow the value to vary during the reconstruction, assuming that $\frac{f_n}{f_p}$ has not been experimentally determined in advance.  Due to the degeneracy between isospin violation and a change in the cross section, allowing $\frac{f_n}{f_p}$ to vary effectively increases the uncertainty in the inferred cross section (see Fig.~\ref{figFnFp} left). The addition of a second detector has the potential to break this degeneracy; however, in practice the astrophysical uncertainties make this impossible. The inclusion of the argon detector greatly improves mass reconstruction, but has a limited effect on reducing the uncertainty in the inferred cross section  (see Fig.~\ref{figFnFp} left and right).  Also, we once again see that there is not much improvement in reconstruction when using two different detector targets compared with doubling the size of the xenon detector. However, it is possible that with the addition of more detectors of different target material, one can at least infer the sign of $\frac{f_n}{f_p}$ \cite{Pato:2011de}.

\begin{figure}[htp]
\centering
\mbox{
\includegraphics[width=80mm]{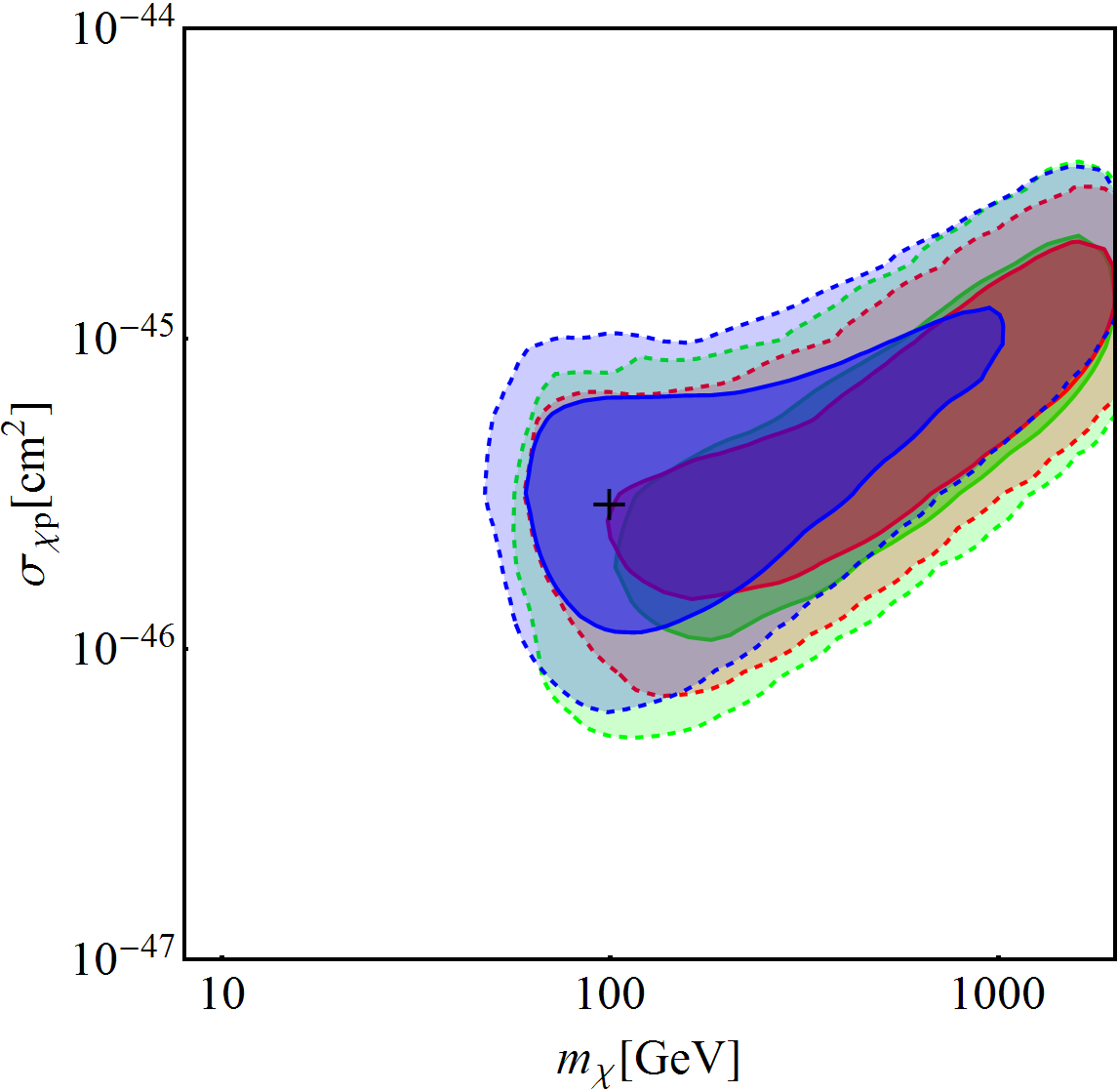}
\includegraphics[width=80mm]{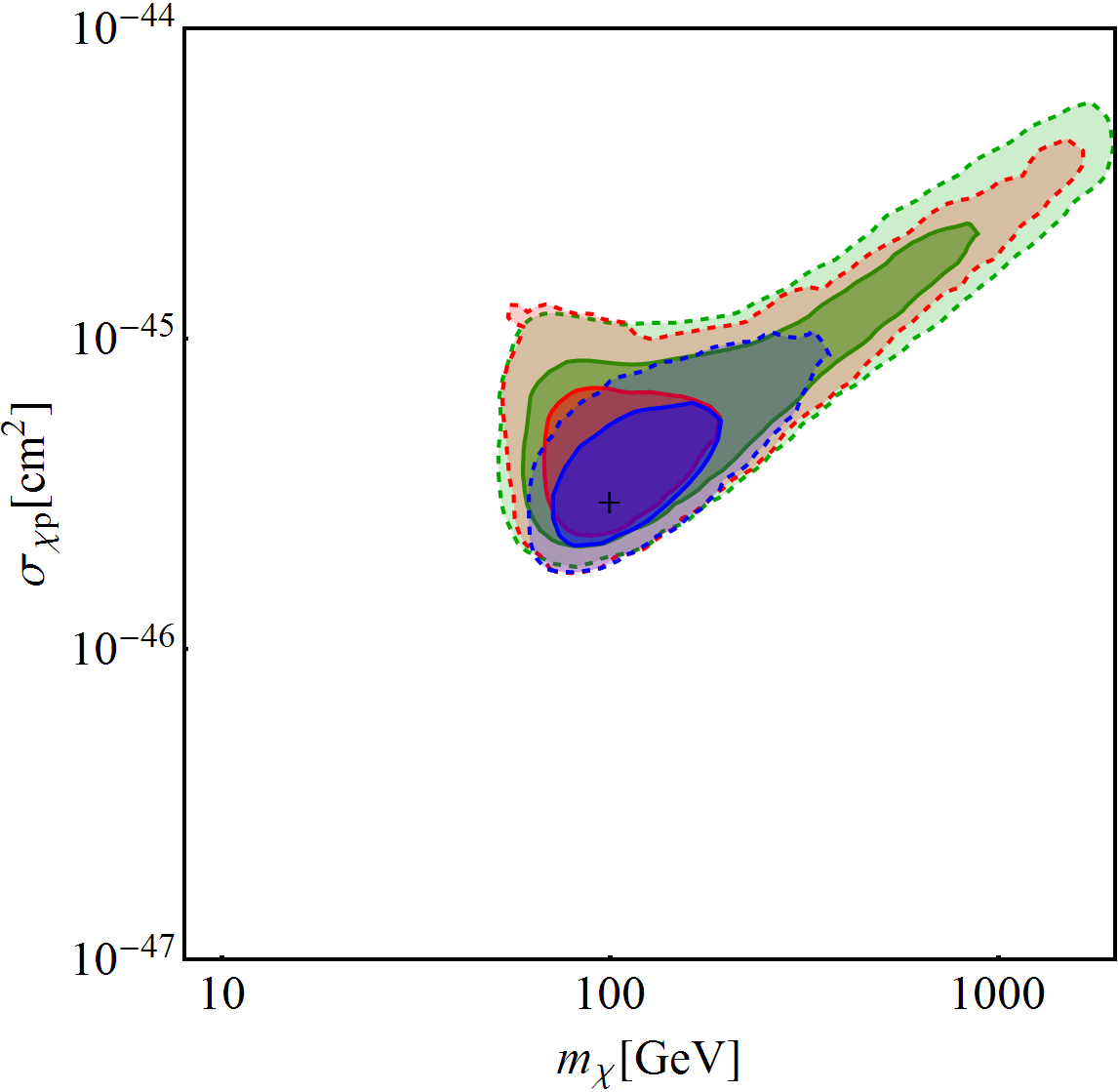}
} 
\caption{One- and two-sigma credible regions of the marginal posterior probabilities for simulations of WIMPs with the same values of $m_\chi$, $\sigma_{\chi p}$ and $\frac{f_n}{f_p}$ as in Fig.~\ref{figFnFp}.   
In both figures the inelastic scattering parameter $\delta$ is allowed to vary during reconstruction. 
Shown are exposures of 10 tonne-years xenon (green), 20 tonne-years xenon (red) and 10 tonne-years xenon plus 20 tonne-years argon (blue). 
(\textit{Left}): $\delta = 0$ keV during simulation, allowed to vary during reconstruction. 
(\textit{Right}): $\delta = 50$ keV during simulation, allowed to vary during reconstruction.}
\label{figDelta1}
\end{figure}

\begin{figure}[htp]
\centering
\mbox{
\includegraphics[width=80mm]{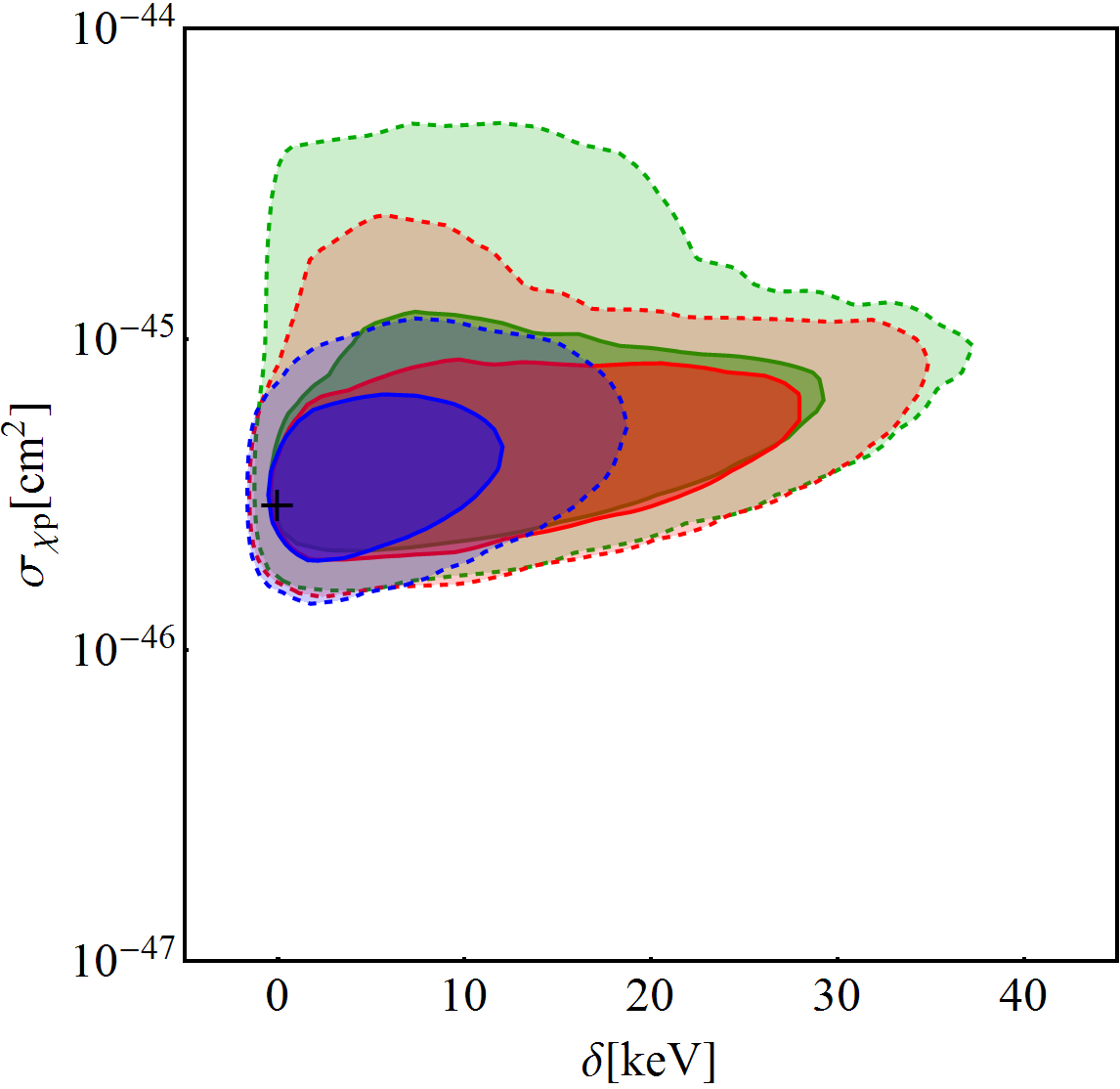}
\includegraphics[width=80mm]{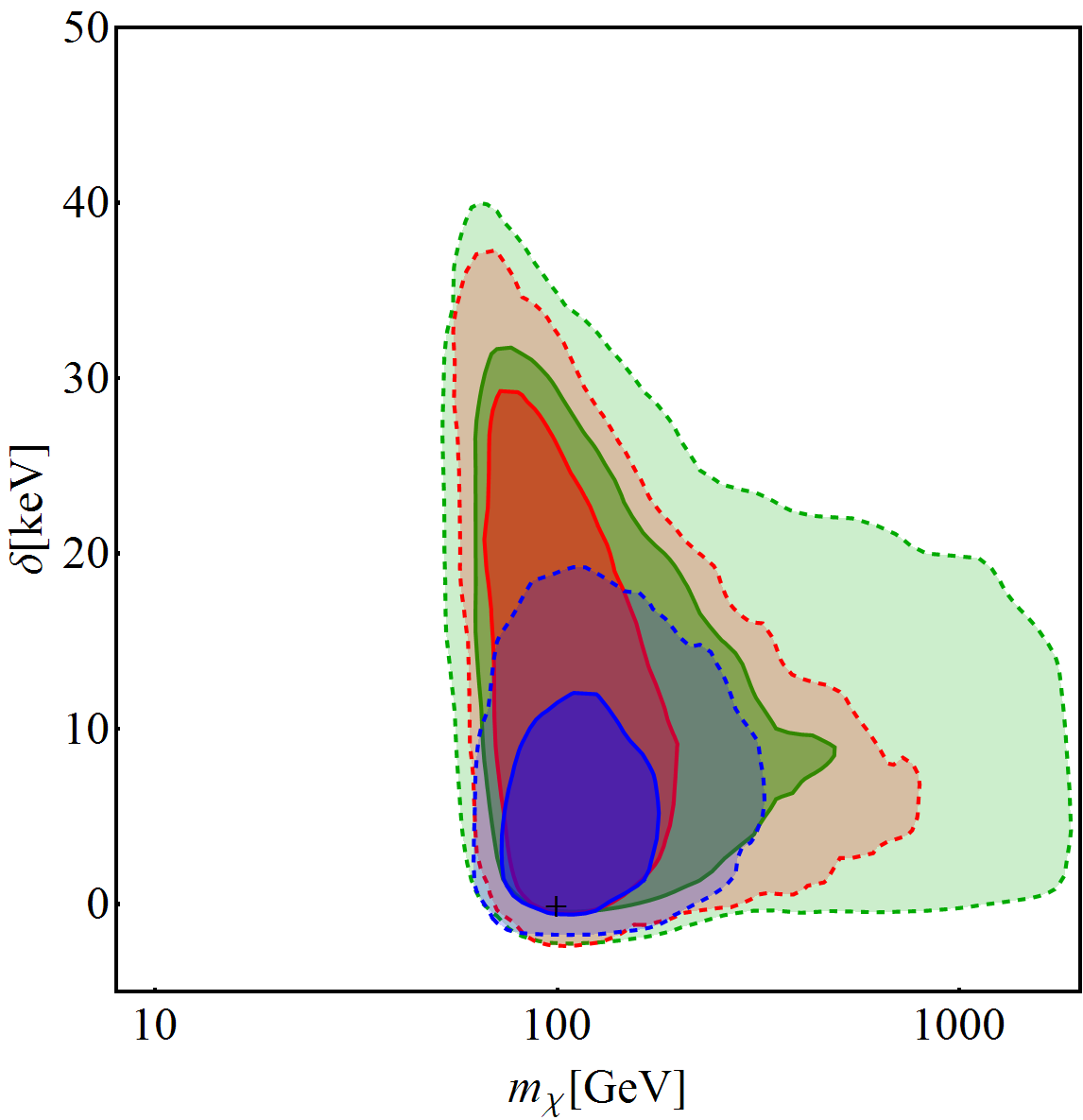}
} 
\caption{One- and two-sigma credible regions of the marginal posterior probabilities for simulations of WIMPs with the same parameters as in Fig.~\ref{figFnFp}. 
In both figures the inelastic scattering parameter is fixed to $\delta = 0$ keV during simulation and allowed to vary during reconstruction. 
Shown are exposures of 10 tonne-years xenon (green), 20 tonne-years xenon (red) and 10 tonne-years xenon plus 20 tonne-years argon (blue). 
(\textit{Left}): Reconstruction in the $\sigma_{\chi p} - \delta$ plane. 
(\textit{Right}): Reconstruction in the $\delta- m_\chi$ plane.}
\label{figDelta2}
\end{figure}

Although a less generic physical possibility, the addition of a non-zero inelastic scattering probability greatly increases the uncertainty in the reconstruction, since the event rate is decreased in this scenario. The event rate is diminished to such an extent that for $\delta = 100$ keV, there are no inelastic events visible for a 100GeV WIMP with $\sigma_{\chi p}=3\times10^{-46}$ cm$^2$. Events are observable for $\delta = 50$ keV, and here the complementarity of the two detectors provides a small improvement in the reconstruction (see Fig.~\ref{figDelta1} left), compared with doubling the xenon exposure. Fixing $\delta=0$ during simulation while allowing it to vary during reconstruction gives a modest increase in the uncertainty in the reconstruction compared to assuming a specific value of $\delta$, shown in Fig.~\ref{figDelta1} right.
The second detector plays a stronger role in the reconstruction of the value of $\delta$, providing a substantially stronger constraint on $\delta$ than obtained by doubling the size of the Xenon component, shown in Fig.~\ref{figDelta2}.

\begin{figure}[htp]
\centering
\mbox{
\includegraphics[width=80mm]{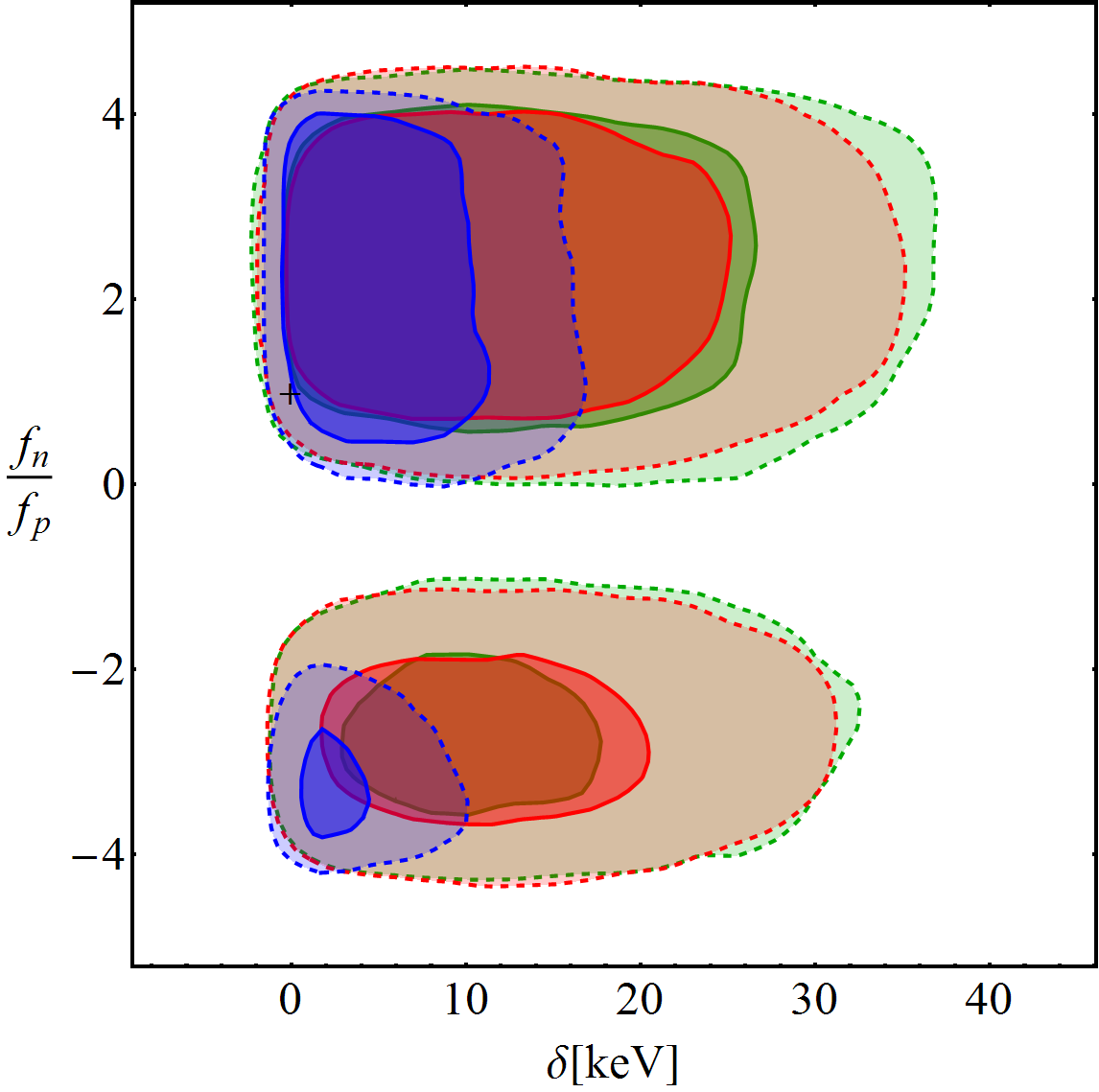}
\includegraphics[width=80mm]{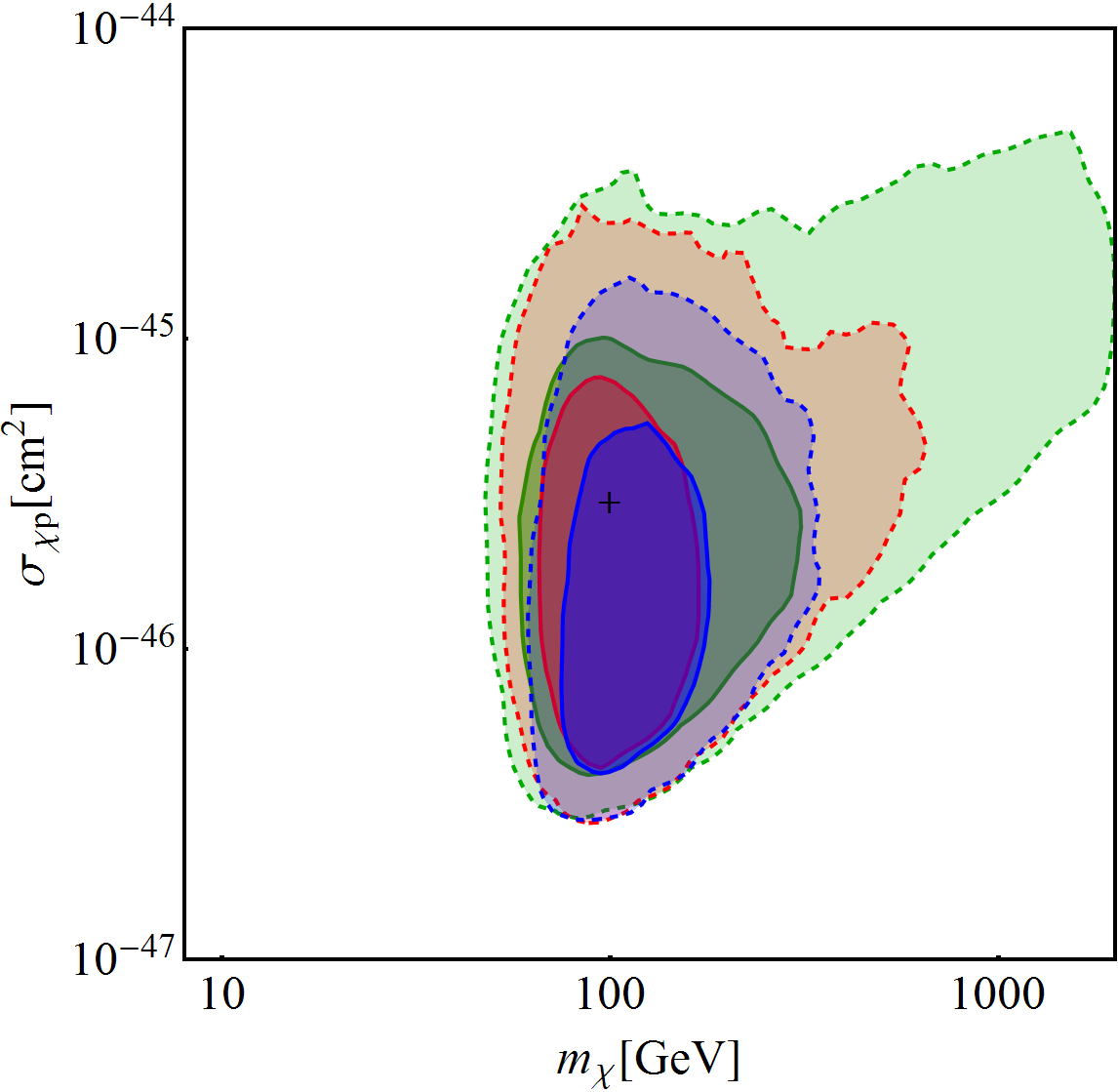}
} 
\caption{One- and two-sigma credible regions of the marginal posterior probabilities for simulations of WIMPs with the same parameters as in Fig.~\ref{figFnFp}. 
In both figures, both $\frac{f_n}{f_p}$ and $\delta$ are fixed to the values in Fig.~\ref{figFnFp}  during simulation and allowed to vary during reconstruction. 
Shown are exposures of 10 tonne-years xenon (green), 20 tonne-years xenon (red) and 10 tonne-years xenon plus 20 tonne-years argon (blue). 
(\textit{Left}):  Reconstruction in the $\frac{f_n}{f_p} - \delta$ plane.
 (\textit{right}): Reconstruction in the $\sigma_{\chi p} - m_\chi$ plane.}
\label{figDeltaAndFnFp}
\end{figure}

Combining these two effects, if we assume neither $f_n/f_p = 1$ nor $\delta = 0$ keV in the reconstruction, then the WIMP properties can only weakly be constrained. Fig.~\ref{figDeltaAndFnFp} left shows that similar to the individual cases, $f_n/f_p$ and $\delta$ are only weakly constrained with individual detectors, while there is a strong improvement in the reconstruction of $\delta$ once the data from the two detectors are combined. Interestingly, large values of $\delta$ seem to prefer positive values of $f_n/f_p$.
Fig.~\ref{figDeltaAndFnFp} right shows that little information can be obtained about the WIMP mass or cross section under these relaxed assumptions. In particular, the reconstruction of the cross section is substantially worse than under the standard assumptions of Fig.~\ref{figReconstruct2} left.

\section{Conclusions}

Given the current understanding of possible WIMP candidates for dark matter, the greatest difficulty in extracting dark matter properties  in direct detection experiments arises from astrophysical uncertainties--in particular the underlying phase space distribution in our halo.  
The existence of two different detector targets, each with similar overall sensitivity but different sorts of systematic uncertainties, will certainly aid in differentiating any claimed signal from possible background, but the question arises as to what extent degeneracies in mass and cross section reconstruction can be further reduced in the event of separate signals in the two detectors.

The DM direct-detection simulation and reconstruction program we have developed addresses this question, in addition to exploring the dominant sources of uncertainty in the expected signal, with some surprising results.  In particular, the complementarity between xenon and argon targets only modestly improves the ability to remove  the degeneracies affecting mass and cross section determinations, and for dark matter particles in excess of around 200 GeV the allowed range in mass-cross section space begins to blow up.  While a number of particle physics parameters produce sub-dominant uncertainties in reconstructing dark matter parameters from an observed signal, the possibility of isospin violation in particular can dramatically increase the uncertainty in derived parameters.  Additional (or a different combination of) detector targets would be needed to try to disentangle the effects of isospin violation from a reduction in cross section.    Improved constraints in halo parameters would assist greatly in reconstruction efforts as well.  

While possible spin-dependent effects in WIMP scattering will further complicate the reconstruction effort, they will also provide another  handle on distinguishing signals from background and exploiting the complementarity of different target nuclei.  Future improvements in our program will determine to what extent the two competing effects will alter the ability to determine WIMP properties based on signals in direct detection experiments. 

\begin{acknowledgments}
J.B.D. thanks the Louisiana Board of Regents and the National Science Foundation for support. F.F. was partially supported for this work by the U.S. DOE under Contract No. DE-FG02-91ER40628. L.M.K, T.D.J. and J.L.N. acknowledge support from the DOE for this work.  We thank Daniel Hunter and Marc Schumann for helpful discussions, Laura Baudis for detailed comments and suggestions on the manuscript, and the Australian National University for hospitality while this work was completed.
\end{acknowledgments}

\bibliography{PhysicsBibtex}

\end{document}